\documentclass[aip,jcp,reprint,graphicx,twocolumn,amsmath,amssymb,groupedaddress]{revtex4-1}

\usepackage{rotating}
\usepackage{dcolumn}               
\usepackage{epsfig}
\usepackage{graphics}

\newcommand{\bra}[1]{\ensuremath{\left\langle{#1}\right\vert}}

\newcommand{\ket}[1]{\ensuremath{\left|{#1}\right\rangle}}

\def\bea{\begin{eqnarray}}
\def\eea{\end{eqnarray}}

\begin{document}

\title{The nature of the low energy band of the Fenna-Matthews-Olson complex: vibronic signatures }
\author{Felipe Caycedo-Soler, Alex W. Chin, Javier Almeida, Susana F. Huelga and Martin B. Plenio}
\affiliation{Institute of Theoretical Physics
Albert-Einstein-Allee 11
D - 89069 Ulm, Germany}
\pacs{82.53.Ps;87.14.E-;87.15.A-,87.15.H-;87.15.M-;87.14.E-}

\begin{abstract} 

Based entirely upon actual experimental observations on electron-phonon coupling, we develop a theoretical framework to
show that the lowest energy band of the Fenna-Matthews-Olson (FMO) complex exhibits observable features due to the quantum nature of the vibrational
manifolds present in its chromophores. The study of linear spectra provides us with the basis to understand
the dynamical features arising from the vibronic structure in non-linear spectra in a progressive
fashion, starting from a microscopic model to finally performing an inhomogeneous average.
We show that the discreteness of the vibronic structure can be witnessed by probing the diagonal peaks of the non-linear spectra
by means of a relative phase shift in the waiting time resolved signal. Moreover, we demonstrate that the photon-echo and non-rephasing paths are sensitive to different harmonics in
the vibrational manifold when static disorder is taken into account. Supported by analytical and numerical calculations, we show that non-diagonal
resonances in the 2D spectra in the waiting time, further capture the discreteness of
vibrations through a modulation of the amplitude without any effect in the signal intrinsic frequency.
This fact generates a signal that is highly sensitive to correlations in the static disorder of the excitonic
energy albeit protected against dephasing due to inhomogeneities of the vibrational ensemble.

\end{abstract}

\date{\today}

\maketitle

\section{Introduction}\label{introduction}
The dynamics of photosynthetic complexes, such as the Fenna-Matthews-Olson (FMO) complex, have received rapidly increasing attention as an example of how nature appears to favor an intermediate noise regime, as suggested by the discovery of experimental evidence for long-lasting coherent oscillations present in signals arising from non-linear spectroscopy protocols \cite{FMOnature,Flemming2005}, and the theoretical discovery that environmental fluctuations play a crucial role in explaining the very high efficiencies for excitation  energy transport in such complexes \cite{Mohseni2008,PlenioHuelga2008,Caruso2009,Chin2010}.  The FMO  complex is a trimer, with individual subunits composed of eight strongly coupled chromophores that serve as a link  between the Chlorosome antennae  and the reaction center of green sulfur bacteria  \cite{Matthews1980,Renger2011}. Chromophores have a spatial extension that supports the existence of intramolecular vibrations  \cite{Czarnecki1997,Mathies1991,Small2000,Small1991}. These vibrations play an important role when the chromophore structures are perturbed due to the electronic charge redistribution upon photon excitation.

Non-linear spectroscopy has proved to be useful in order to unveil the dynamics involved in excitonic transfer of light harvesting complexes, due to the fact that it is sensitive to excitonic quantum superpositions, i.e., excitonic coherences.  Two dimensional (2D) non-linear spectroscopy can resolve the third-order polarization of the electronic system, from the heterodyne detected spectrally resolved signal $S(t_1,t_2,\omega_{3})$, arising from photo-excitation of three consecutive pulses with wave vectors ${\bf k_1}, {\bf k_2}$ and ${\bf k_3}$, separated by time intervals $t_1$ and $t_2$ \cite{Mukamel}.  Experimentally, it is customary to Fourier transform the time  $t_1$ dimension to yield  $S(\omega_{1},t_2,\omega_{3})$ in order to generate a   2D spectra  parametrized by the waiting time $t_2$.  This spectra allows inquiring into the dynamics of excitation transfer in photosynthetic light harvesting complexes.

The appreciation of the role of discrete vibrational modes on the excitation transfer process has evolved over time \cite{Cina2012}, and  additional signatures of their presence in the 2D signal will contribute to a thorough understanding.  The main purpose of this paper is to show that the experimental evidence points towards the presence of a discrete vibronic structure in the lowest energy band of FMO, and that the dynamical contribution of these vibronic modes can be extracted from the information available through non-linear spectra protocols.  \\
Particularly, the multiplicity of states resolved in the low energy band in the FMO  has  been the subject of intense debate. It has not been possible to date, to uniquely assign its nature to electronic contributions. In this manuscript we propose a model for the lowest energy band in the FMO that includes explicitly the most significant resonances of its environment spectral density, which is consistent with the available experimental evidence achieved through hole burning \cite{Small1991,Ratsep1998,Small2000}, fluorescence line-narrowing \cite{Small1991,Amerongen2000},  and accumulated photon echo \cite{Louwe1997}. We  explore the possibility to detect the coupling of intra-molecular  low  energy vibrations of chromophores, to exciton dynamics, and shed light on the nature of the lowest energy band of  the FMO (to be referred as B825 from its absorbance maximum). We show that the experimental evidence can be explained with the existence of a representative vibronic resonance with energy $\simeq 36$ cm$^{-1}$ and, beyond current suppositions that neglect higher order contributions, its overtone.   Interestingly, this resonance has also  been specifically addressed in resonant Raman spectra of the special pair BChl{\it a} in purple bacteria reaction centers \cite{Mathies1994}, inferred through the Stokes shift  in LHC-II protein complexes Chl{\it a} pigment  of higher plants \cite{Renger1997}, and exists within the energy range expected for low-frequency vibrations of helical structures in proteins \cite{Chou1983a,Chou1983b}. This strongly suggests that its origin are either the absorbing chromophores or a ubiquitous protein feature such as the $\alpha$-helices both of which, are therefore not specific  to the FMO structure.
 
In order to introduce our model we comment on the outcome of different experimental examinations in the B825, and qualitatively reproduce their results with the assumption of the presence of a single excitonic state, a discrete vibronic dynamical contribution, and its overtone. The dynamical contribution of this vibronic resonance is studied through the phase difference among signals obtained in the neighborhood of an electronic diagonal peak in the 2D spectra,  which allows  to distinguish the fundamental from higher vibrational  harmonics.  By unraveling the underlying  vibrational dynamics in the 2D spectroscopy time domain, we study  the characteristics of the beating signal that arise due to this electron-phonon interaction (the phonon term will be used for the intramolecular vibrations of our interest).  Accordingly,  we explore the dynamics of non-diagonal contributions in the 2D spectra, to show that a modulation of the amplitude in the excitonic coherence signal  arises due to these low frequency vibrational modes. We obtain an  analytical expression for the excitonic coherence that explains this beating pattern  as a result of vibrational wave-packet motion. We highlight that the coherent oscillations are extremely sensitive to the degree of correlation in the static disorder of neighboring chromophores and propose a physical picture involving the spatially extended $\alpha$-helices of the protein scaffold, that is able to explain the observed long lifetime for excitonic coherence.  On the other hand, the modulation of the signal amplitude due to vibrational wave packet motion does not alter the intrinsic frequency of oscillations of the polarization signal in 2D spectra and therefore, produces no dephasing due to inhomogeneities of the vibrational manifold. The latter effect allows us to explain recent experimental results where, even though structural modification in the FMO chromophores were accomplished, the dephasing rate of the electronic coherence in such a complex remained unaltered \cite{Engel2011b}.
  
 \subsection{Brief history of the FMO low energy band B825} Due to the energetic difference between the first exciton and higher lying excitonic states, the lowest energy band in the FMO complex has become the optical transition in which to study and understand,  the extremely long lifetime of excitonic coherences found in this photosynthetic complex \cite{FMOnature}. Nevertheless, the results obtained with different experimental techniques have not reached a  consensus on the nature of this energy band. 
 
 The first hole burning and fluorescence line-narrowing experiments performed in FMO \cite{Small1991}, showed two components  in the B825, whose absorption differed by $\Delta\omega\approx$ 30 cm$^{-1}$. The possibility that  these frequencies were zero-phonon and phonon side bands was considered but  ruled out at the time, because the energy difference between emission and absorption maxima in the B825 was $\Delta E\simeq$40 cm$^{-1}$, and could not be reconciled with the  calculated Huang-Rhys  factor $s=0.3$ from measured absorbance difference, that resulted in a Stokes shift $2s\omega=2(0.3)30\approx 20$ cm$^{-1}\ne \Delta E$. Rather, these levels were regarded as excitonic states arising from delocalized excitonic components of the FMO trimer \cite{Small1991}, a nature  corroborated by theoretical calculations  of the whole FMO trimer    \cite{Pearlstein1992}  that suggested the existence of equivalent BChl's 7  (using the numbering scheme of Ref.\cite{Matthews1980})  on different units  as the origin of the three excitonic states on the B825. 
   Even though linear dichroism confirmed delocalization on B825 due to the broad linewidth  in triplet-minus-singlet spectra  \cite{vanGrondelle1994}, the linear dichroic absorption with respect to a continuous microwave field  showed that the angle among the triplet and $Q_y$ transition dipoles in the B825 is constant throughout the band \cite{Louwe1997c}, and close to that observed in single BChl{\it a} \cite{Vrieze1995}.  Since the triplet and singlet orientations are sensitive to excitonic delocalization, the aforementioned equivalence among the B825 and single BChl{\it a}, suggested that such a band was composed of localized excitations within a single BChl{\it a}. This last turn of events was even further corroborated  when the simultaneous fitting of absorption spectra, triplet-minus-singlet and linear dichroic absorbance, concluded that the B825 was composed of localized excitons on the BChl 3  \cite{Louwe1997b}. Shortly afterwards, Stark hole burning studies \cite{Ratsep1998}  compared the purple bacteria B800 band  (composed of almost completely localized excitonic states) with the B825 to provide further support  that in the B825, the inhomogeneous broadening among equivalent BChls over the trimer generates excitons localized on single monomers of the FMO. The low energy band was successfully fitted with three gaussians, that were thought to represent the contributions of the localized lowest exciton states of each subunit of the FMO  trimer \cite{Ratsep1999}. The view that the contributions in B825 would come from  energetic inequivalence arising from structural heterogeneity  was also supported by the absence of excitonic dipole moment redistribution,  probed by polarized hole burning for this energy band \cite{Small2000}.
 In that work,  two important results  were exposed. First, a greater homogeneous linewidth was found with greater energies within the B825, and confirmed the trend found by accumulated photon echo, of higher dephasing rates as the involved pulses were blue-shifted \cite{Louwe1997}.  Second,  a very rich structure of satellite low energy holes regarded as intramolecular vibrations (with the biggest components having energies of 36 and 72 cm$^{-1}$) in the B825 was first observed.   
 
 The trend of increased linewidth at higher energy in the B825 was attributed then, to cascade excitation F\"orster energy transfer among individual chromophores on different subunits. However, Monte Carlo simulations of the Hamiltonian  under the hypothesis of these three different chromophores being  BChl 3  \cite{Small2002}, were not able to simultaneously fit the lineshape and the  gradual variation of the homogeneous linewidth and  dephasing times observed.  Almost concurrently with hole-burning experiments, the technique of fluorescence line-narrowing   \cite{Amerongen2000} successfully determined the phonon spectral density and confirmed the  hole-burning experiments  \cite{Small2000} about the existence of  discrete phonons with frequencies 36  and 70 cm$^{-1}$  moderately interacting with excitons. The possibility  that multiple dynamical components on this band could arise from these modes, was first suggested. 
 
  Difference fluorescence line narrowing \cite{Ratsep2007}  provided further evidence of three contributions on the B825,  and suggested once again, the possibility of exciton delocalization in the lowest exciton state since new simulations of the B800 in purple bacteria harvesting structure, showed a delocalization length across 2-3 pigments \cite{Silbey2006}. Recently, molecular dynamics simulations  \cite{Strumpfer2011} showed an appreciable narrowing of the of site energies in a trimer compared to a monomer. This narrowing renders infeasible that static structural inhomogeneities in the trimer are the cause of the B825 multiple dynamic contributions. 
 
At present, the consensus view is that of a single excitonic state in the low energy band  to describe the FMO optical spectra. This view was corroborated  with more detailed calculations of the FMO electronic Hamiltonian  \cite{Rengereview}, that have shown increasing success in fitting optical spectra, when linewidth assignments of such transitions involved a fit of the  spectral density coarse grained envelope with a global Huang-Rhys factor $s\simeq 0.5$.  However,  no model has been able explain  simultaneously the facts that within the B825, 1) there is a single electronic excitation  where  at least two representative resonances appear, 2) a marked trend of greater homogeneous linewidths (or dephasing rates) while probing higher energies was found. Beyond these points (highlighted in a recent review \cite{vangrondelle2010}), the fact that 3) a very rich intramolecular  vibrational structure is observed but never explicitly accounted for. This  suggests the lack of a complete understanding of  the low energy band of FMO.

\section{The model}
Our immediate purpose is to explain the possible nature of the multiple resonances found with increasing linewidth for blue-shifted probing of  the B825. At the 
same time, the model must be consistent with the approximate  30 cm$^{-1}$ absorption energy difference among these transitions, with the observed Stokes shift and with the fluorescent line narrowing spectral density.   The dipole moment of such transitions must not exhibit variations in polarization while scanning this  energy band.
Let us examine these issues in detail. \\
The most recent Huang-Rhys factor $s=0.5$ describing the global electron-vibration interaction,  is better able to explain the  Stoke's shift of 40 cm$^{-1}$ observed, occurring mainly due to the 36 cm$^{-1}$ mode, i.e., $2s\omega=2(0.5)36\approx 40$ cm$^{-1}$. The whole phonon wing Huang-Rhys factor is an upper bound for the interaction strength among this mode and electronic excitations. In particular, hole burning spectra \cite{Small2000} assigned Huang-Rhys factor $s=0.12$ for the most representative resonances, namely,  resonances having energies of 36 and 72 cm$^{-1}$. Fluorescence line narrowing \cite{Amerongen2000} reported  for both these resonances, a Huang-Rhys of $s=0.01$, on a procedure that might overestimate the weight of the phonon wing.  As the authors explicitly point out, it is heuristically assumed that individual modes just refine details over the phonon wing, and therefore minimally contribute to build up the maximum of the spectral density function. 
The result of either model differ beyond the purely practical aspect of a fitting procedure, since a discrete contribution may lead to coherent evolution that will result in crucial features to be studied below.  Therefore, if the coupling of this mode is upper bounded by  the global Huang-Rhys factor $s=0.5$, and lower bounded by the reported  hole-burning Huang-Rhys value $s=0.12$,  a perturbative analysis is dubious.  Moreover, the vibronic contribution does not represent a redistribution of dipole moment,  allowing a single polarization throughout the B825. Hence, we proceed with the explicit inclusion of this mode. 

 The Hamiltonian for electron-vibration interaction reads
\begin{equation}\label{Heph}
H_0=\sum_k\left(\frac{\omega_0}{2}+\sqrt{s_k}\omega(b_k+b_k^+)\right)\sigma_z+\omega b_k^+b_k
\end{equation}
where $\sigma_i$ and  $b_k^{(+)}$ are usual spin 1/2 Pauli matrices, and annihilation (creation) operators of the $k$th boson field,   the electron-vibration coupling strength $\sqrt{s_k}\omega_k$ is given in terms of the phonon frequency $\omega_k$ and Huang-Rhys factor $s_k$. \\
In the present section, our aim is to consistently describe the homogeneous broadening trend as the probing laser field is blue shifted when vibrations are explicitly included. To that end we consider a single vibration and the  laser interaction with the charge distribution of the chromophore under the rotating wave approximation (RWA) in a frame rotating at the laser frequency, we obtain a hamiltonian:
\begin{equation}
H=H_0+(\Delta-\omega_0)\sigma_z+\frac{\Omega}{2}\sigma_x
\end{equation}
where  $\Delta=\omega_0-\omega_L$ and $\Omega$ are laser detuning from the  energy of the zero phonon line (ZPL) electronic transition and Rabi frequency, respectively.

Now, the broadening of the resonances is of utmost importance in the historical discussion and is required for the analysis of spectra,  as  it influences the interaction of the system with the environment. The homogeneous   linewidth of the resonances will primarily depend on the different processes inducing decoherence beyond the ensemble average. Accumulated photon echo \cite{Louwe1997} fitted the dephasing rate in 1.8-50 K range  with the expression:
\begin{equation}\label{fit}
\frac{1}{T_2}=\frac{\Gamma}{2}+\gamma_zT^{1.3}+\gamma\frac{1}{e^{\delta E/kT}-1}
\end{equation}
 for the homogeneous dephasing time $T_2$, and  highlight the main processes involved:  population relaxation, pure dephasing from spectral diffusion promoted by the surrounding protein, and a thermally induced activation process. Given that the FMO complexes used in optical experiments lack the RC, the first term in eq.(\ref{fit})  is  lower bounded by the fluorescent relaxation. The second term, is attributed to dynamics of   tunneling among energy minima of different structural configurations in the protein, and characterized by a rate that obeys a power law temperature dependence $\propto T^{1.3}$, typical of glassy hosts. The third terms involves  thermally activated transitions not specified in \cite{Louwe1997}, that  accounts here for the thermal equilibration of the vibrational manifold. Note that eq.(\ref{fit}) only describes thermal relaxation, and therefore is best suited to the red edge of the B825. A fit to the red edge photon echo signal was accomplished with free parameters $\{T_2,\gamma_z,\gamma,\delta E\}=\{420$ ps,0.09 ns$^{-1}/K^{1.3},\, 23.6$ ns$^{-1},15\, $cm$^{-1}\}$.   In our model the most prominent thermally activated process occurs among states whose energies involve  boson quanta of energy 36 cm$^{-1}$ which restricts $\delta E=36$cm$^{-1}$. A good agreement among both sets of parameters is accomplished for a constant  $\gamma=70.09 $ns$^{-1}$=2.36 cm$^{-1}$ (results not shown).

In a first model, these mechanisms can be described under the Born-Markov approximation with super-operators $\mathcal{L}_i(\rho)=L_i\rho L_i^+-\frac{1}{2}( L_i^+L_i\rho+\rho L_i^+L_i)$ in a Lindblad-type master equation ($\hbar=1$):
\begin{equation}\label{meq}
\partial_t\rho=-i[H,\rho]+\sum_i\mathcal{L}_i(\rho)
\end{equation}
with $L_1=\sqrt{\Gamma}\sigma^-$ for population relaxation ($\sigma^{- (+)}$ are lowering (rising) Pauli spin ladder operators) and $L_2=\sqrt{2(\bar{n}+1)\gamma} b$, $L_3=\sqrt{2\bar{n}\gamma}b^+$ describing thermalization to temperature $T=1/\beta k$ of the mode having an equilibrium average quanta $\bar{n}=(e^{\beta\omega}-1)^{-1}$ \cite{Rivas2010}. The factor 2 in this latter pair is introduced since the dephasing produced is half the rate of relaxation/absorption processes. The protein configuration tunneling induced dephasing  is accounted by the  operator $L_4=\sqrt{\gamma_{z}T^{1.3}}\sigma_z$.  The magnitude of rates $\Gamma$, $\gamma$ and $\gamma_z$ are hence provided  from the dephasing rate  obtained through accumulated photon echo technique \cite{Louwe1997}.

\section{Results}

\subsection{Linear spectra}

\begin{figure*}
\includegraphics[width=.8\columnwidth]{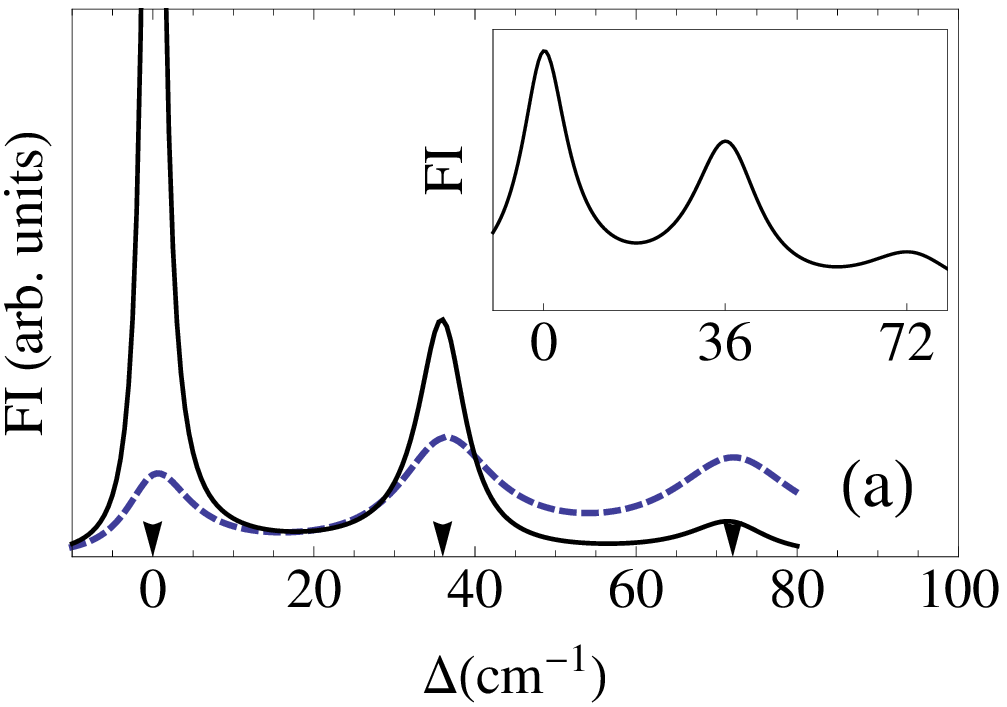}
\includegraphics[width=.8\columnwidth]{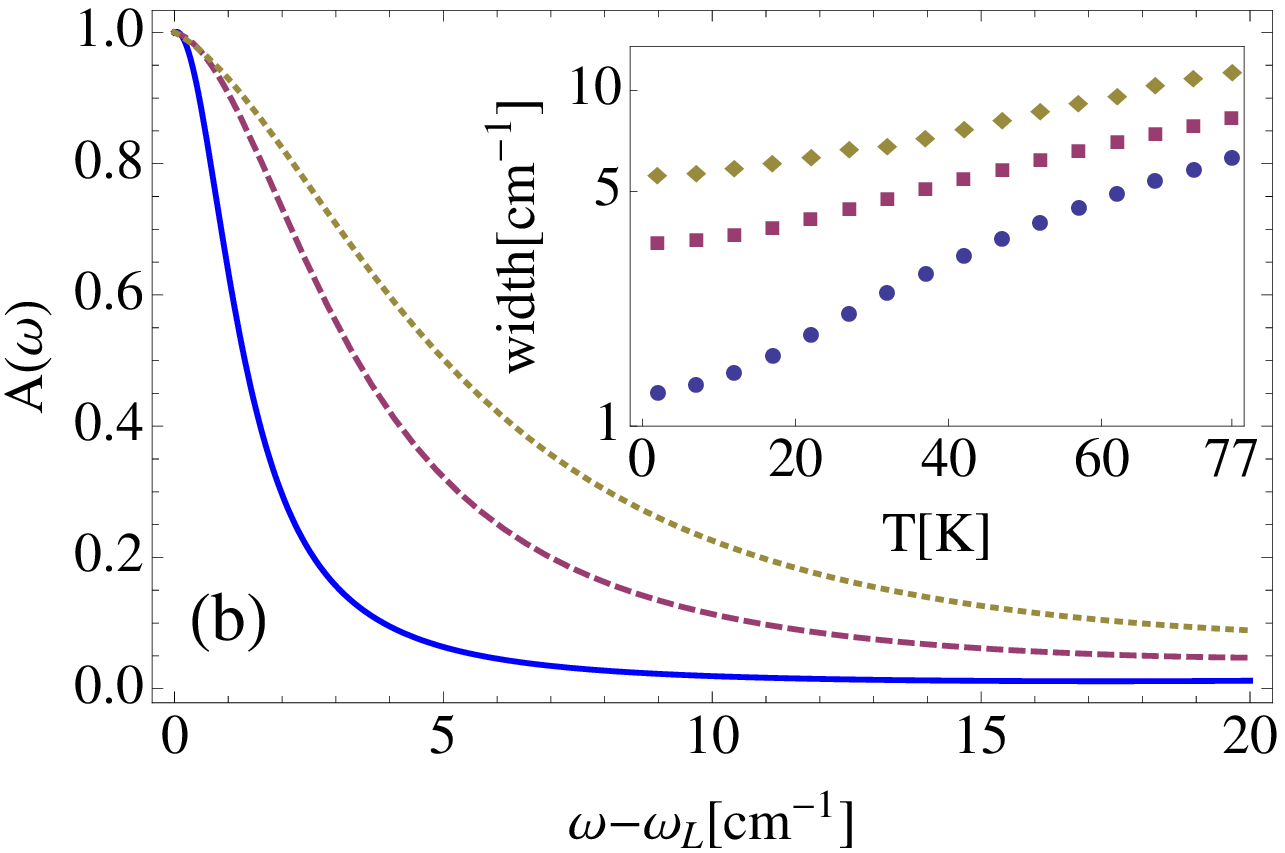}
\caption{In (a) fluorescence intensity  from solution of stationary state in eq.(\ref{meq}). Dashed and continuous correpond to Huang-Rhys factors $s=0.5$ and $s=0.12$, at T=4K, drawn with the same scale. The arrows point the frequency of the zero-phonon (0 cm$^{-1}$), phono side-band (36 cm$^{-1}$), and overtone (72 cm$^{-1}$). The inset shows the fluorescence intensity at T=77K, $s=0.12$.  In (b) $A(\omega)$ is normalized for the three resonances and correspond in continuous, dashed and dotted, to the $\Delta=0$, 36 and 72 cm$^{-1}$ resonances, that have a FWHM  of  $\{\gamma_0,\gamma_1,\gamma_2\}=\{1.6, 3.5,5.6\}$ cm$^{-1}$ at 4 K, from a fit with lorentzian functions.  Inset (b), shows the variation of the lorentzian widths with temperature for the ZPL ($\gamma_0$, circles), sideband at 36 ($\gamma_1$, boxes) and excited harmonic 72 cm$^{-1}$ ($\gamma_2$, diamonds),  that yields to values of $\{\gamma_0,\gamma_1,\gamma_2\}=\{6.2, 8.1, 11.2\}$cm$^{-1}$ at 77 K. If otherwise not stated $\Gamma=7.9\times 10^{-2}$ cm$^{-1}$, $\gamma=2.36$ cm$^{-1}$, $\gamma_{z0}=3\times 10^{-3}$ cm$^{-1}$,  $s=0.12$ , $\omega=36$ cm$^{-1}$,  $\Omega= 10^{-3}$ cm$^{-1}$.}\label{fig0}
\end{figure*}

The fluorescence intensity  in the line narrowing signal is proportional to the excited state population in the stationary state of the chromophore $\langle \sigma^+\sigma^-\rangle_{t\rightarrow \infty}$, obtained from eq.(\ref{meq}), and  presented in Fig.\ref{fig0}(a). First, note that with both the upper and the lower bounds of the Huang-Rhys factor, three fluorescent transitions  (highlighted by arrows) can be resolved: the zero-phonon line, the phonon side band, and the overtone, which coincides with the 70-72 cm$^{-1}$ resonance reported in \cite{Small2000}.  It should be  noted that there is no need to include another vibration  70-72 cm$^{-1}$ since a single mode and its harmonic lead to the most representative resonances at $\simeq 36$ and $\simeq 72$ cm$^{-1}$. Increasing the Huang-Rhys factor to $s=0.5$ results in a higher overlap between levels having different boson quanta in the excited and ground electronic states, and therefore a redistribution of fluorescence intensity  increases the contribution of higher harmonics. It should be noted that even  a moderate Huang-Rhys factor $s=0.12$ results in three dynamical contributions at $T=$77 K, as shown in the inset of Fig.\ref{fig0}(a). At this temperature, the homogeneous phonon side band is commensurate with the zero-phonon line, and therefore, even in the presence of ensemble inhomogeneities, both transitions should be of importance in the description of the FMO in 2D spectroscopy ($T=$77 K, Ref.\cite{FMOnature}).
 
Having captured these three dynamical contributions, we proceed with the analysis of the individual linewidths. The hole burning experiments, sample the absorbance $A(\omega)$ proportional to the Fourier transform (FT) of a two-time correlation function, explicitly:
\begin{equation}
A(\omega)\propto \lim_{t\rightarrow\infty}\int_0^\infty \langle\sigma^-(t)\sigma^+(t+\tau)\rangle e^{i(\omega_L-\omega)\tau}d\tau , 
\end{equation}

\noindent which is presented in Fig.\ref{fig0}(b)  (calculated with the quantum regression theorem according to the Lindblad form master equation eq.(\ref{meq})) for the relevant burning frequencies. In this figure, it is shown that higher energy on the burning beam generates greater linewidths of the holes in the absorption spectrum.  With this model, the  linewidths, say $\gamma_0$, $\gamma_1$, $\gamma_2$, of the ZPL, the phonon side-band, and overtone, respectively, show a difference that persists for the whole  range of temperatures measured as shown in the inset of this figure. This is in complete agreement with photon echo experiments (1.8-50 K, \cite{Louwe1997}).  The broadening can be understood from transitions  involving $n$-th  vibrational states, that become more populated with increasing temperature and whose thermalization rate is proportional to the excitation number $\langle \mathcal{L}_{2(3)}\ket{n}\bra{n}\rangle\propto n$. The model also captures two important facts found in hole burning and photon echo experiments. Firstly, the greatest difference between linewidths occurs at low temperatures  with a trend to converge  as temperature is increased  \cite{Small2000}. Secondly, before convergence is reached to equal linewidths, the three components are widening in the same fashion \cite{Louwe1997}. This behavior illustrates the prevalence of different mechanisms depending on the temperature.  At low $T$,  spontaneous emission and thermalization decay are the key mechanisms, the former being  independent of temperature. At intermediate temperatures, dephasing induced  by protein conformations is equal in any of the transitions and represents the trend of the linewidths to equalize. At high $T$, the onset of thermalization as the dominating dephasing mechanism induces equal increment in all linewidths with temperature. In practice, the reddest photon echo signal according to eq.(\ref{fit}) from accumulated photon data  \cite{Louwe1997}, extrapolates to a dephasing rate of 3.3 cm$^{-1}$. Based in our theoretical results, this rate is  $\approx\gamma_0/2= 6.2/2$ cm$^{-1}$ (see the inset Fig.\ref{fig0}(b)  at T=77 K). The difference of the data with our ZPL linewidth can be traced back to mixing of echoes originating from the ZPL and the side-band within the ensemble.

At this stage, we conclude that by inclusion of a vibrational mode with frequency of 36 cm$^{-1}$ and an Huang-Rhys factor $s\ge 0.12$, we are able to capture three dynamical contributions with equal polarization, whose position in absorption spectra and linewidths trend follow the experimental observation in the lowest energy band of the FMO. Without inclusion of a such mode, the absorptive response showing three dynamical contributions below T=10 K \cite{Small2000} cannot be explained relying solely on an inhomogeneously broadened single excitation electronic contribution. 

Additional information is available in the 2D non-linear spectra that might better characterize this low energy vibration, and further support (or reject) its role in the additional dynamical component on the lowest energy band of the FMO.

\subsection{Non-linear 2D spectroscopy}\subsubsection{Background} In 2D spectroscopy, the heterodyne detected signal $S(\omega_1,t_2,\omega_{3})$ allows straightforward comparison with the FT of the theoretical analogue 
\begin{eqnarray}
S(\omega_{1},t_2,\omega_{3})&=&\mbox{FT[2 Im}\left(\sum_{i=1}^4{R_i(t_1,t_2,t_3)}\right)](\omega_1,t_2,\omega_3)\nonumber\\
 \label{signal}
\end{eqnarray}
for a two level  electronical system  with energy difference $\omega_{eg}$ among excited and ground states $\ket{e},\ket{g}$ respectively. In this case, the response functions $R_2$ and $R_3$ ($R_1$ and $R_4$)  usually termed as rephasing  (non-rephasing) paths, allow calculation of signal  in the direction of observation ${\bf k_s}=-{\bf k_1}+{\bf k_2}+{\bf k_3}$   (${\bf k_s}={\bf k_1}-{\bf k_2}+{\bf k_3}$) surviving the rotating wave approximation.  In order to present the data in a single quadrant, the FT for the non-rephasing and rephasing signals have conjugate variables $\{-\omega_1,\omega_3\}$ and $\{\omega_1,\omega_3\}$, respectively. In detail, 
 
 \begin{equation}R_n=\vert \mu^4\vert e^{\pm i\omega_{eg} (t_1\mp t_3)}e^{f_n(t_3,t_2,t_1)}\end{equation} 
 
\noindent where upper (lower) signs correspond to rephasing (non-rephasing)  terms, respectively. In the above expression, $\mu$ is the electronic transition dipole moment  and the function $f_n(t_3,t_2,t_1)$ depends on the lineshape  function $g(t)$ evaluated in combinations of these three time intervals \cite{Mukamel}. The lineshape function of a Brownian oscillator 

\begin{equation}\label{linewidth}g(t)=i s\omega t+s(i\sin \omega t+\coth (\beta\hbar\omega/2)[1-\cos\omega t])\end{equation}  describes the contribution to the response function from a discrete vibration with frequency $\omega$  \cite{Mukamel}.  In this particular case, the  function $g(t)$ generates through $f_n(t)$ \cite{Mancal2010}, an increase of   the excitonic frequency by the reorganization energy $s\omega$, $\omega_{eg}\rightarrow\omega_{eg}+s\omega$, produces a global reduction of the signal by the Debye-Waller factor squared, $e^{-2s\coth(\beta\omega/2)}$, and has  additional time dependent terms of the form $e^{s(i \sin(\omega t)-\coth(\beta\omega/2)\cos(\omega t))}$. The Fourier transforms required for the 2D plots  are performed here analytically from a Taylor series expansion  $e^{Os} \approx 1+Os+(O s)^2+\cdots$. The exponential function has a rapid convergence on this expansion for $s\coth(\beta\omega/2)\ll1$,  partially fulfilled for the  Huang-Rhys factors of interest at the actual temperature of experiments ($0.12\le s\le 0.5, \omega=36$ cm$^{-1}$, T=77 K; $s\coth(\beta\omega/2)=0.37-1.53$). A   first order expansion of this kind  was developed  in  Ref. \cite{Mancal2010}. In the following we further develop the expansion to second order, to study the possibility of resolving the higher harmonics (conspicuous in the absorption and fluorescence intensity) and develop a set of features able to better characterize vibronic resonances in the 2D spectra.

The decay of the time resolved heterodyne signal  in the impulsive limit, is equivalent to the linewidths of spectrally resolved observation under continuous-wave illumination. Henceforth, the effects of the bath in the 2D spectra will be introduced  in the aforementioned expansion for terms  proportional to $e^{\pm i \omega t_1}$, $e^{\pm i\omega t_3}$, $e^{\pm 2i\omega t_1}$, $e^{\pm 2i\omega t_3}$, with the Fourier transform  argument $\omega_{1(3)}+i\gamma_1$ on the former two, and $\omega_{1(3)}+i\gamma_2$ on the latter pair, to include the homogeneous broadening for side-bands and overtone. The terms lacking of any of these factors, are Fourier transformed with an argument $\omega_{1(3)}+i\gamma_0$ modeling the width of the zero phonon line. The addition of these imaginary contributions in the Fourier transform argument is consistent with a decay to the stationary state in the waiting time domain due to the substitution $e^{i\omega t_2}\rightarrow e^{i\omega t_2}e^{-\gamma_1 t_2}$ and $e^{2i\omega t_2}\rightarrow e^{2i\omega t_2}e^{-\gamma_2 t_2}$. 
\begin{figure*}
\includegraphics[width=1\columnwidth]{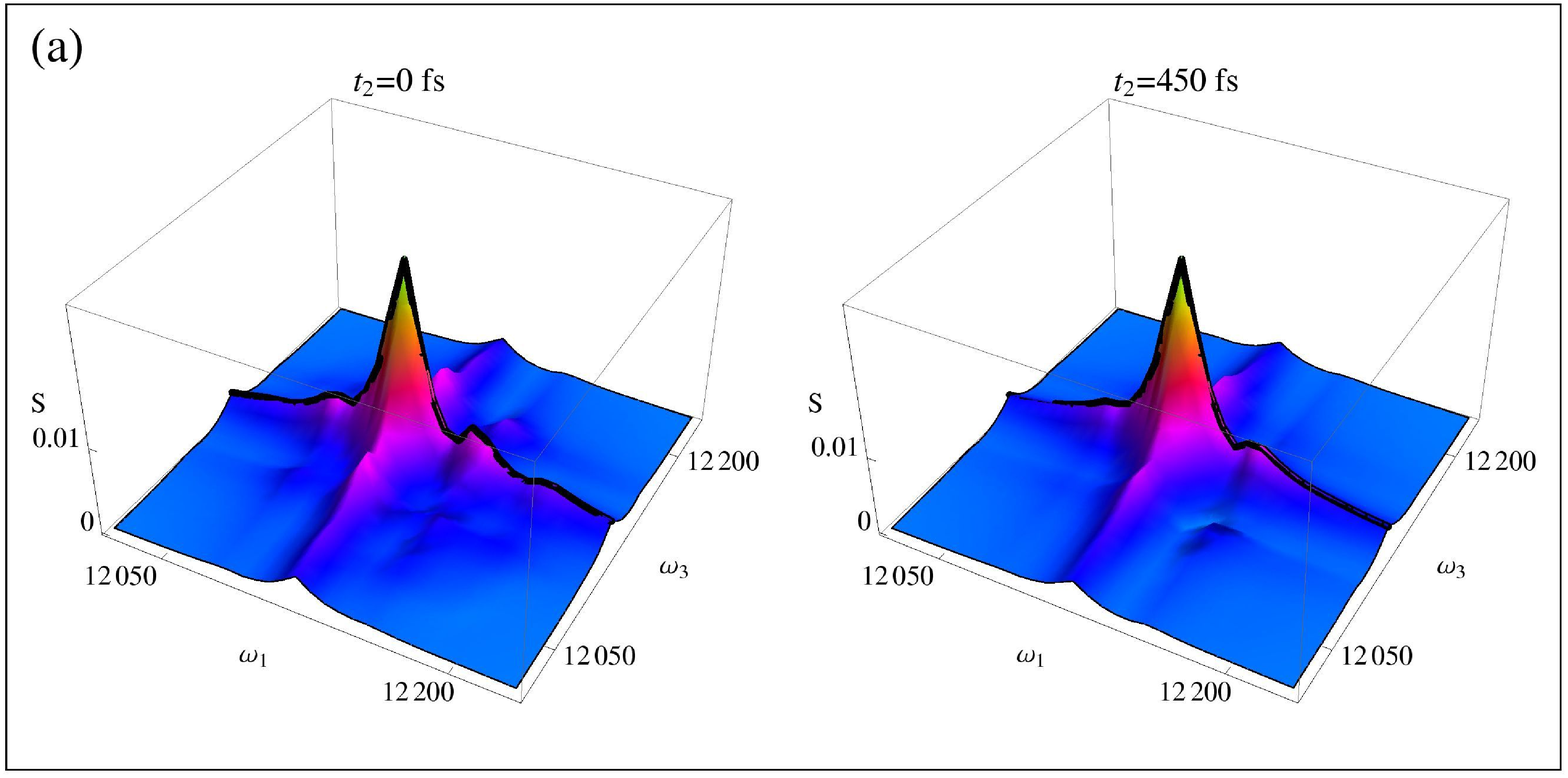}
\includegraphics[width=1\columnwidth]{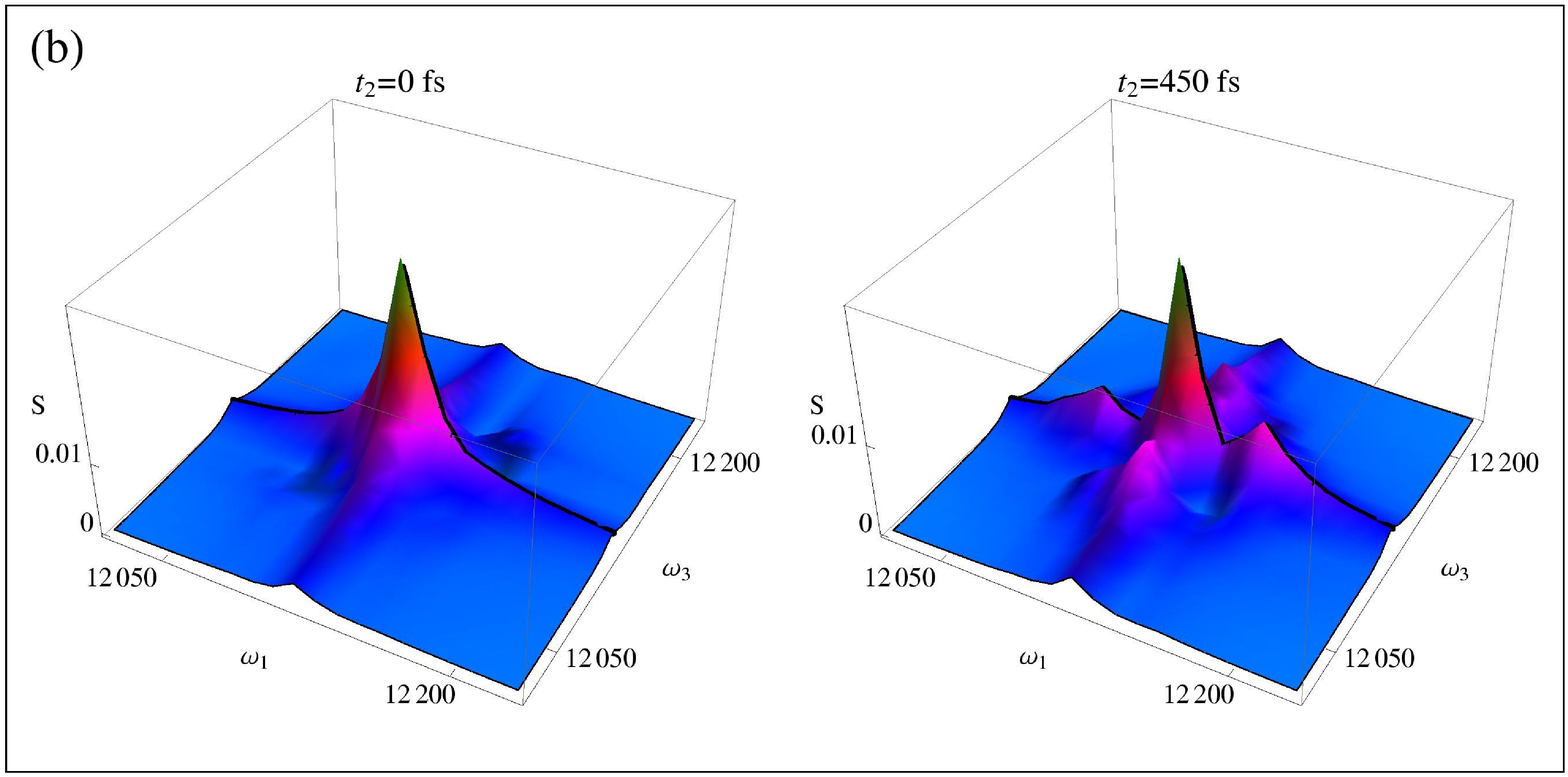}
\includegraphics[width=.7\columnwidth]{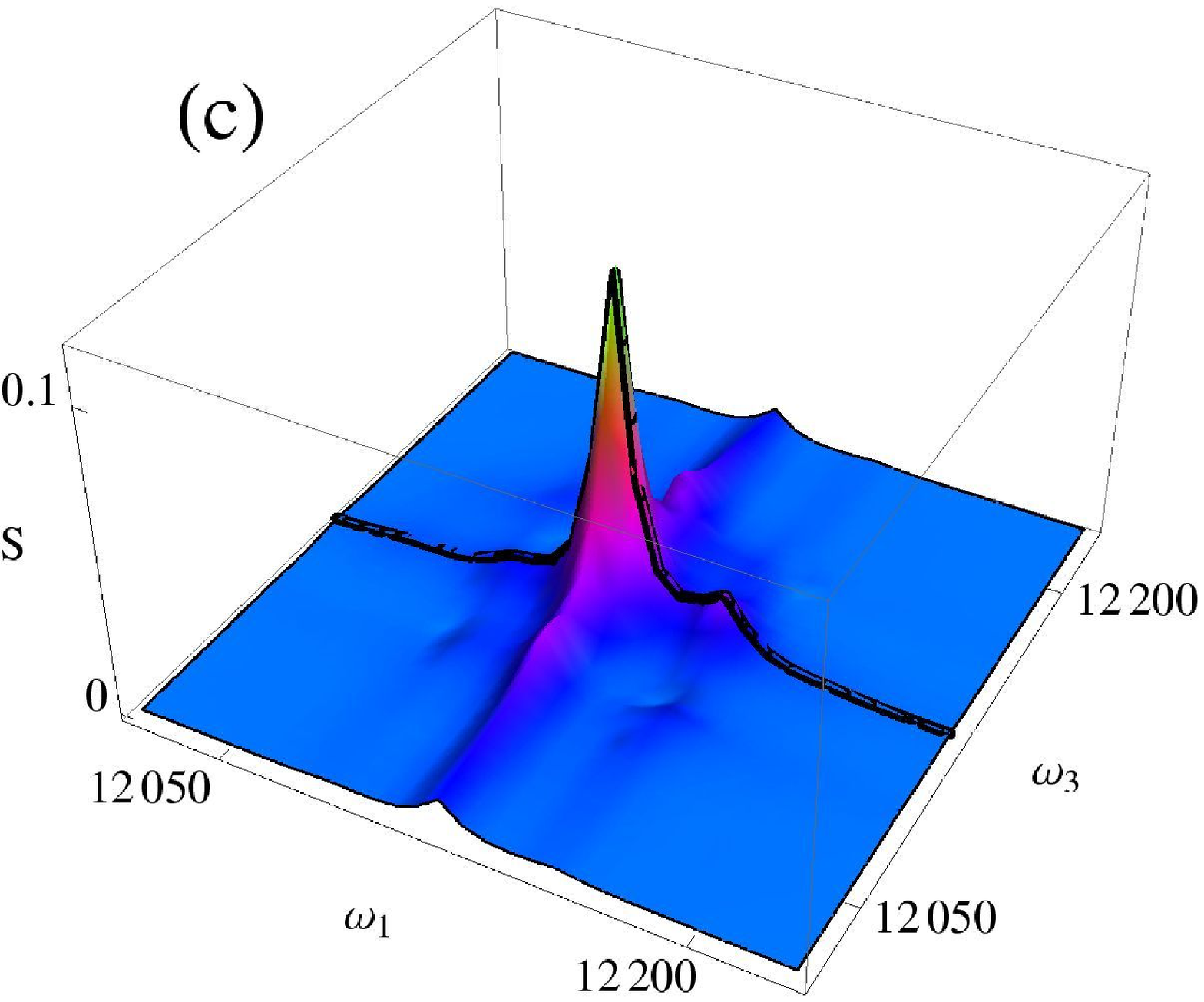}
\includegraphics[width=.7\columnwidth]{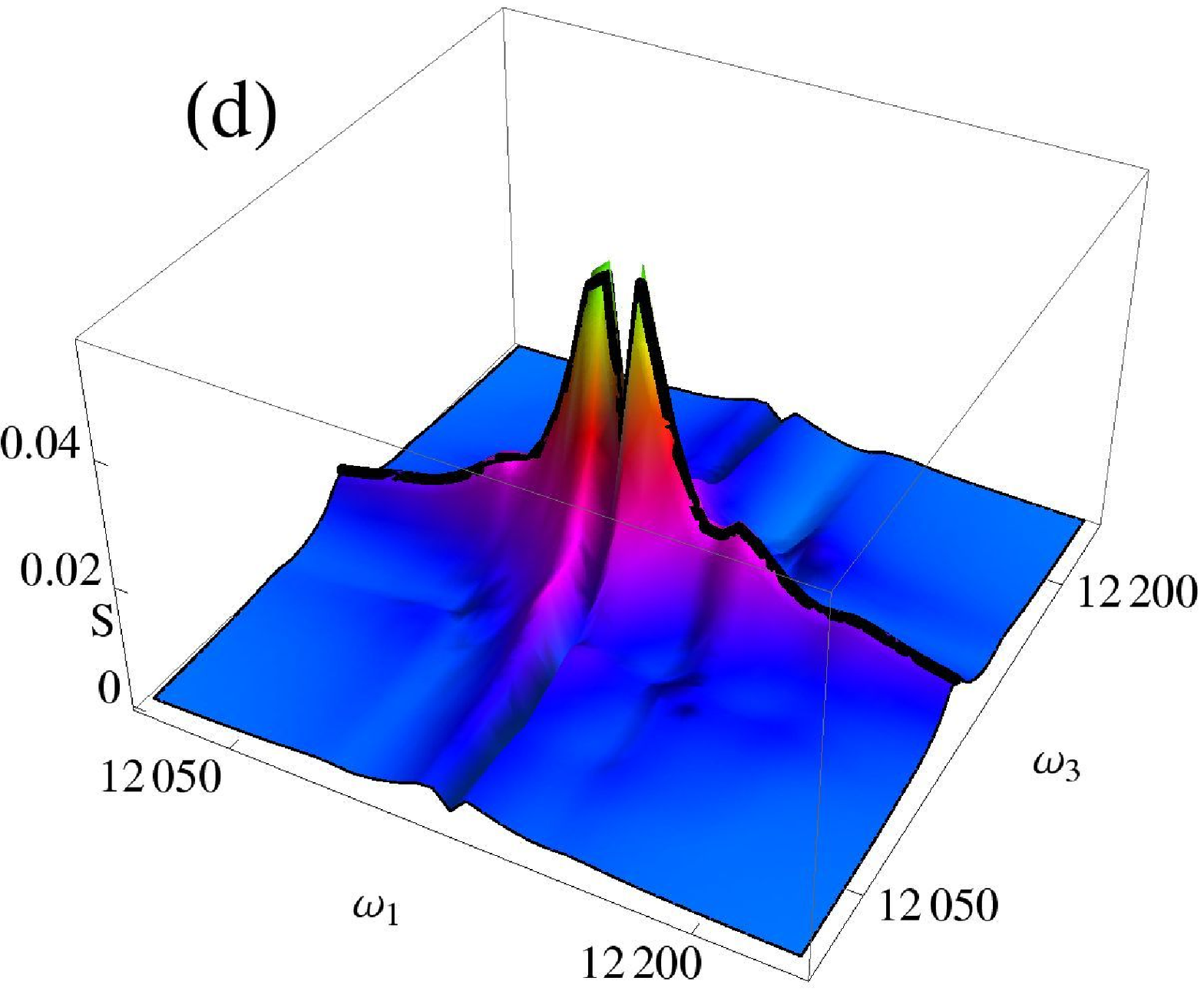}
\caption{ In (a) and (b) we present the second order expansion for rephasing and non-rephasing paths, respectively. (c) the addition of rephasing and non-rephasing contributions,  while (d) subtraction FT$[R_2+R_3-(R_1+R_4)]$. In plots (a)-(b) an arcsinh scale has been used in order to discuss the side-band features, while (c)-(d) use a linear scale for purposes of comparison and  $t_2=450$ fs.  In all plots the zero phonon emission line is shown in continuous line to guide the eye.   Exciton 1 energy $\tilde\omega_{eg}=\omega_{eg}+\sum_i s_i\omega_i=$12121 cm$^{-1}$  \cite{Engel2011}, vibrational mode energy $\omega=36$ cm$^{-1}$, and Huang-Rhys factor $s=0.12$. }\label{fig1}
\end{figure*}

\subsubsection{Vibronic features in 2D spectra: Diagonal peaks} Genuine vibronic sidebands in 2D spectra are challenging to discriminate from  the electronic  contributions. Fortunately the exciton 1, $\ket{\psi_1}$, in the  FMO, has an energy separation $> 150$cm$^{-1}$ from all  other excitonic states which  allows its  spectral resolution from the other electronic contributions  \cite{Engel2011,FMOnature}. The 2D spectra can be used  in order to characterize the most salient features of the vibronic spectra in this lowest energy band, such as the resonances found at 36 and 72 cm$^{-1}$.

\begin{figure*}
 \includegraphics[width=1.7\columnwidth]{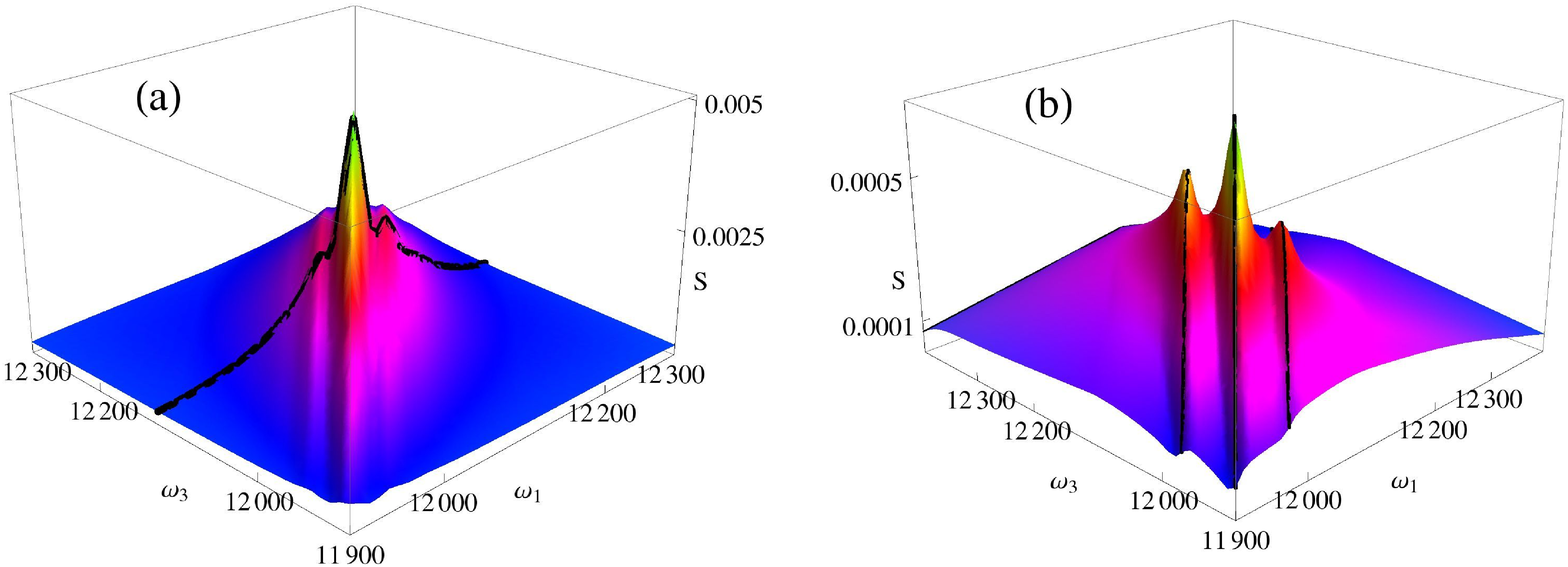}\\
\includegraphics[width=.65\columnwidth]{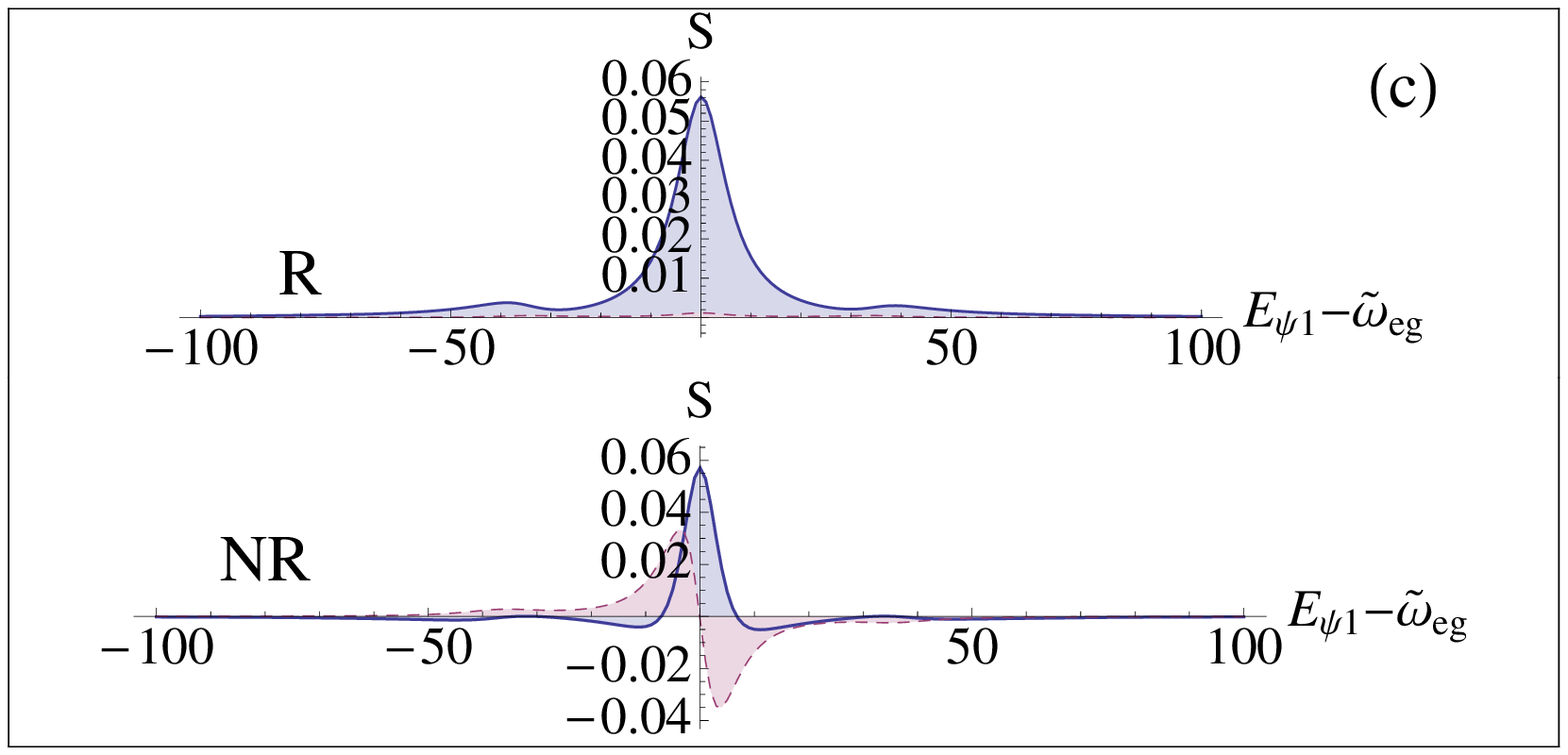}
\includegraphics[width=.65\columnwidth]{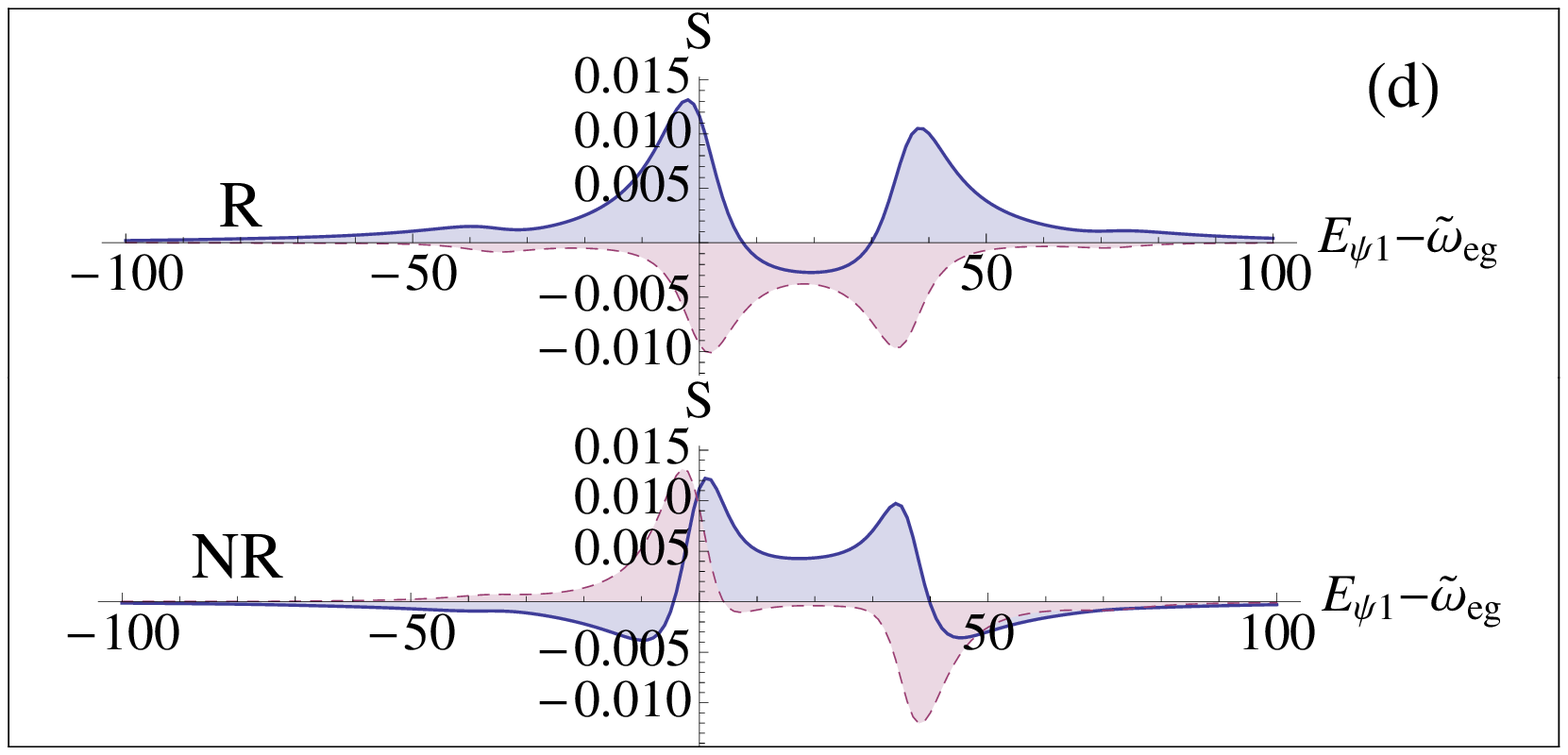}
\includegraphics[width=.65\columnwidth]{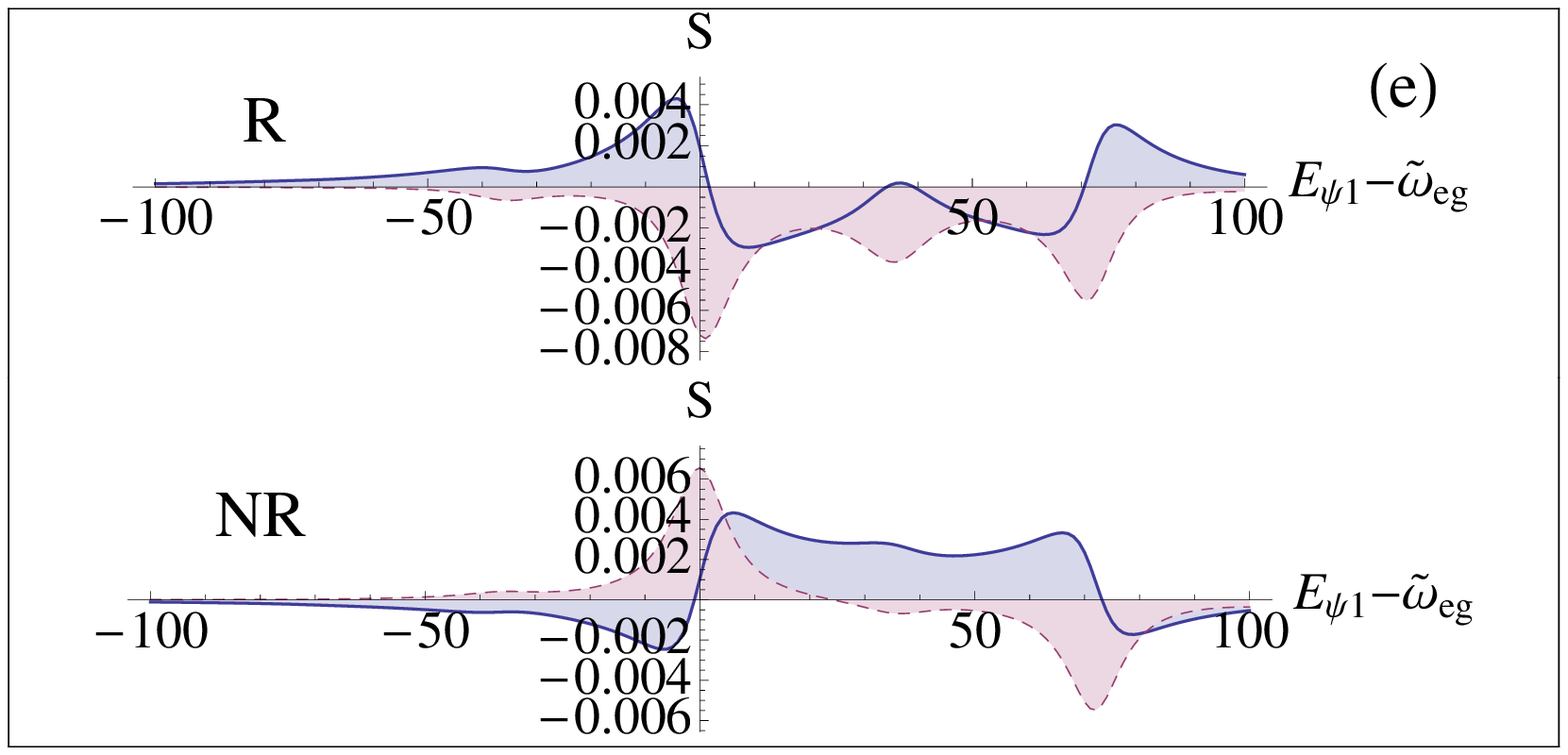}\\
\includegraphics[width=1.7\columnwidth]{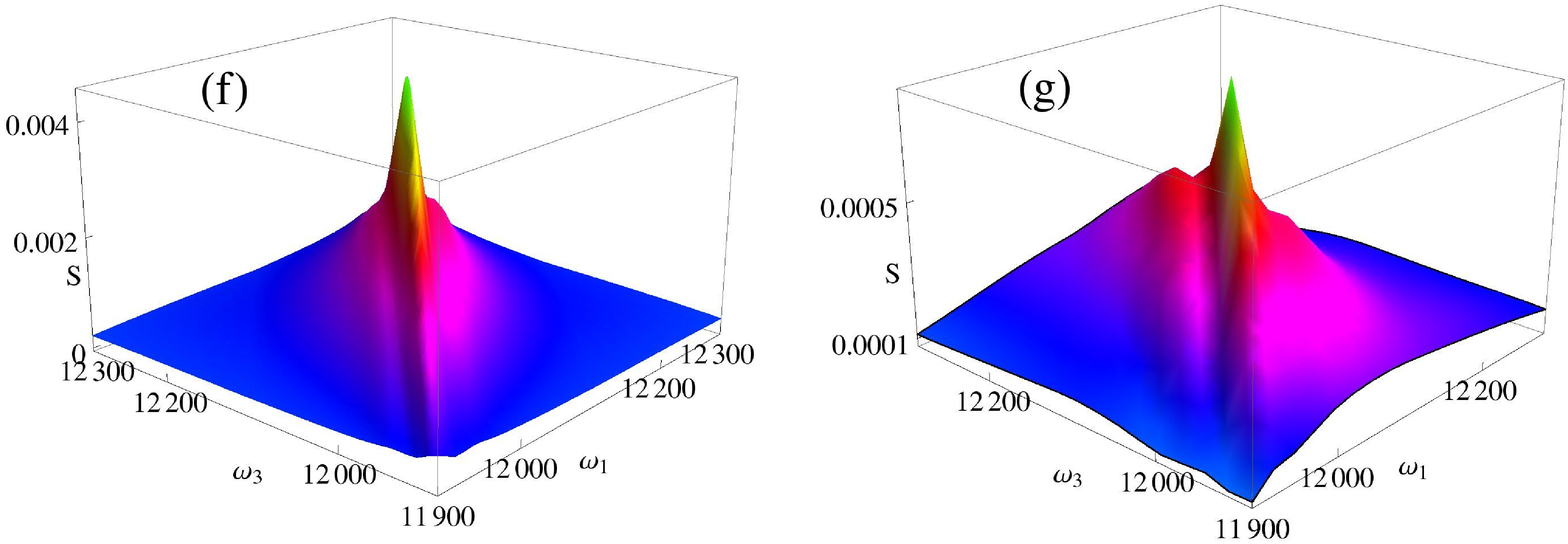}
\caption{In (a) and (b), rephasing and non-rephasing 2D spectra of exciton 1 diagonal peak of FMO complex, broadened by excitonic energy inhomogeneities. In (c), (d) and (e) are presented the real (continuous, blue online) and imaginary (dashed, red online) contributions to the inhomogeneous signal at points $(\omega_1=\omega_3=\tilde{\omega}_{eg})$, $(\omega_1=\tilde{\omega}_{eg}+\omega,\omega_3=\tilde{\omega}_{eg})$ and $(\omega_1=\tilde{\omega}_{eg}+2\omega,\omega_3=\tilde{\omega}_{eg})$ respectively, for complexes whose energy differ $E_{\psi_1}-\tilde{\omega}_{eg}$  from the ensemble average as explained in the text.  In (f) and (g), rephasing and non-rephasing 2D spectra of exciton 1 diagonal peak of FMO complex, broadened by both excitonic and vibrational mode energy inhomogeneities. $\sigma_E=102$ cm$^{-1}$, $\sigma_\omega=10$cm$^{-1}$, $t_2=2$ ps.}\label{fig2}
\end{figure*}

\begin{figure}
\begin{minipage}{1\columnwidth}
\setlength{\unitlength}{0.75mm}
\begin{picture}(85,45)
\put(-10,-6){\line(0,1){54}}
\put(100,-6){\line(0,1){54}}
\put(-10,-6){\line(1,0){110}}
\put(-10,48){\line(1,0){110}}
\put(-10,-6){\line(0,-1){105}}
\put(100,-6){\line(0,-1){105}}
\put(-10,-60){\line(1,0){110}}
\put(-10,-111){\line(1,0){110}}

\put(-9,44){(a)}
\put(2,5){$_{\ket{g^n}\bra{g^n}}$}
\put(5,7){\line(0,1){30}}
\put(10,7){\line(0,1){30}}
\put(5,20){\ldots}
\put(5,14){\ldots}
\put(6,17){$_{t_1}$}
\put(6,23){$_{t_2}$}
\put(6,29){$_{t_3}$}
\put(5,26){\ldots}
\put(5,32){\ldots}
\put(2.5,11){\begin{rotate}{30}$\leadsto$\end{rotate}}
\put(11.7,25){\begin{rotate}{30}$\leadsto$\end{rotate}}
\put(14.4,18){\begin{rotate}{150}$\leadsto$\end{rotate}}
\put(5,33){\begin{rotate}{150}$\leadsto$\end{rotate}}
\put(2,40){$_{\ket{g^{n}}\bra{g^{n}}}$}
\put(-6,17){$_{\ket{e^{n+1}}}$}
\put(12,23){$_{\bra{e^{n}}}$}
\put(-.5,10.5){$ _{k_1}$}
\put(12,16){$ _{-k_2}$}
\put(14,30){$_{k_3}$}
\put(-3,36.5){$_{-k_s}$}
\put(5,-5){$R_1$}

\put(27,5){$_{\ket{g^n}\bra{g^n}}$}
\put(30,7){\line(0,1){30}}
\put(36,7){\line(0,1){30}}
\put(30,20){\ldots}
\put(30,14){\ldots}
\put(32,17){$_{t_1}$}
\put(32,23){$_{t_2}$}
\put(32,29){$_{t_3}$}
\put(30,26){\ldots}
\put(30,32){\ldots}
\put(27,17){\begin{rotate}{30}$\leadsto$\end{rotate}}
\put(36.7,25){\begin{rotate}{30}$\leadsto$\end{rotate}}
\put(40.4,11){\begin{rotate}{150}$\leadsto$\end{rotate}}
\put(30,33){\begin{rotate}{150}$\leadsto$\end{rotate}}
\put(27,40){$_{\ket{g^n}\bra{g^n}}$}
\put(23,23){$_{\ket{e^n}}$}
\put(37,16.5){$_{\bra{e^{n+1}}}$}
\put(23.5,17.5){$ _{k_2}$}
\put(37,8.5){$ _{-k_1}$}
\put(39,30){$_{k_3}$}
\put(22,36.5){$_{-k_s}$}
\put(30,-5){$R_2$}

\put(52,5){$_{\ket{g^n}\bra{g^n}}$}
\put(55,7){\line(0,1){30}}
\put(61,7){\line(0,1){30}}
\put(55,20){\ldots}
\put(55,14){\ldots}
\put(57,17){$_{t_1}$}
\put(57,23){$_{t_2}$}
\put(57,29){$_{t_3}$}
\put(55,26){\ldots}
\put(55,32){\ldots}
\put(52,23.5){\begin{rotate}{30}$\leadsto$\end{rotate}}
\put(61.7,19){\begin{rotate}{30}$\leadsto$\end{rotate}}
\put(65.4,13){\begin{rotate}{150}$\leadsto$\end{rotate}}
\put(55,33){\begin{rotate}{150}$\leadsto$\end{rotate}}
\put(52,40){$_{\ket{g^n}\bra{g^n}}$}
\put(48,30){$_{\ket{e^n}}$}
\put(62,17.5){$_{\bra{e^{n+1}}}$}
\put(48.5,23.5){$ _{k_3}$}
\put(62,10){$ _{-k_1}$}
\put(64,24){$_{k_2}$}
\put(47,36.5){$_{-k_s}$}
\put(55,-5){$R_3$}

\put(82,5){$_{\ket{g^n}\bra{g^n}}$}
\put(85,7){\line(0,1){30}}
\put(91,7){\line(0,1){30}}
\put(85,20){\ldots}
\put(85,14){\ldots}
\put(87,17){$_{t_1}$}
\put(87,23){$_{t_2}$}
\put(87,29){$_{t_3}$}
\put(85,26){\ldots}
\put(85,32){\ldots}
\put(82.5,11){\begin{rotate}{30}$\leadsto$\end{rotate}}
\put(81.7,23){\begin{rotate}{30}$\leadsto$\end{rotate}}
\put(85.4,21){\begin{rotate}{150}$\leadsto$\end{rotate}}
\put(85,33){\begin{rotate}{150}$\leadsto$\end{rotate}}
\put(82,40){$_{\ket{g^n}\bra{g^n}}$}
\put(74.3,17.5){$_{\ket{e^{n+1}}}$}
\put(78,30.5){$_{\ket{e^n}}$}
\put(78.5,10.5){$ _{k_1}$}
\put(75,21.5){$ _{-k_2}$}
\put(77.5,24.5){$_{k_3}$}
\put(77,36.5){$_{-k_s}$}
\put(85,-5){$R_4$}
\put(-9,-13){(b)}
\put(-9,-67){(c)}
\end{picture}
\end{minipage}

\begin{center}

\hspace{0.3 cm}\begin{minipage}{.42\columnwidth}
\includegraphics[width=\columnwidth]{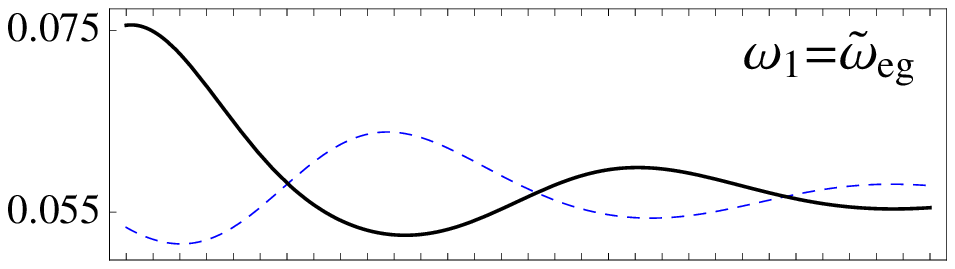}
\includegraphics[width=\columnwidth]{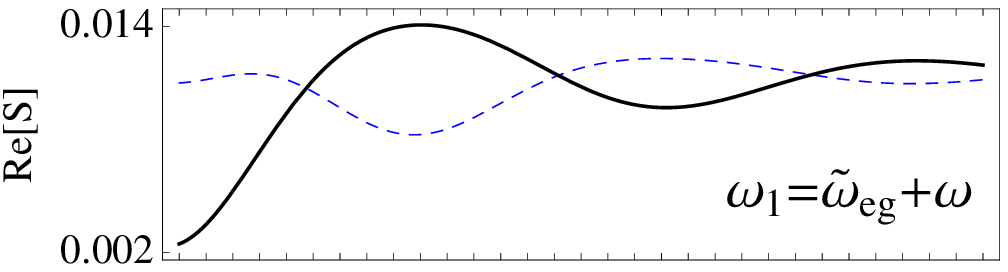}
\hspace{0. cm}\includegraphics[width=\columnwidth]{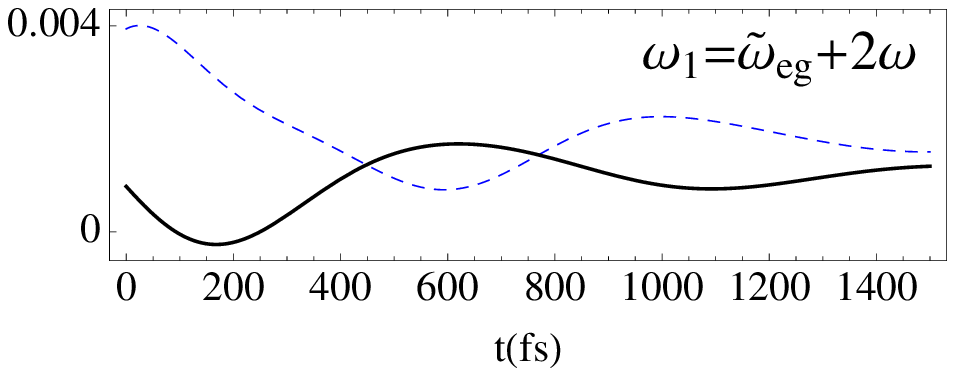}
\end{minipage}
\begin{minipage}{.47\columnwidth}
\hspace{.24 cm}\includegraphics[width=.85\columnwidth]{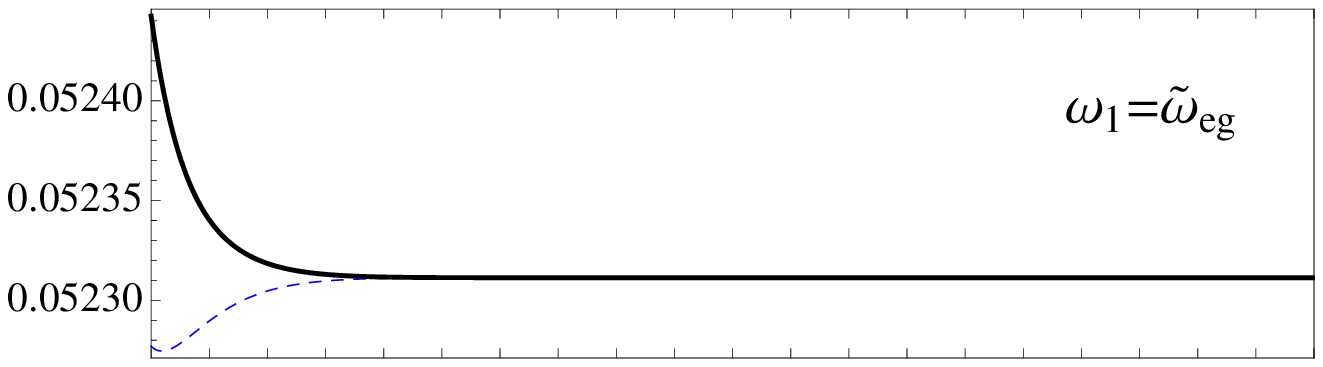}
\includegraphics[width=.92\columnwidth]{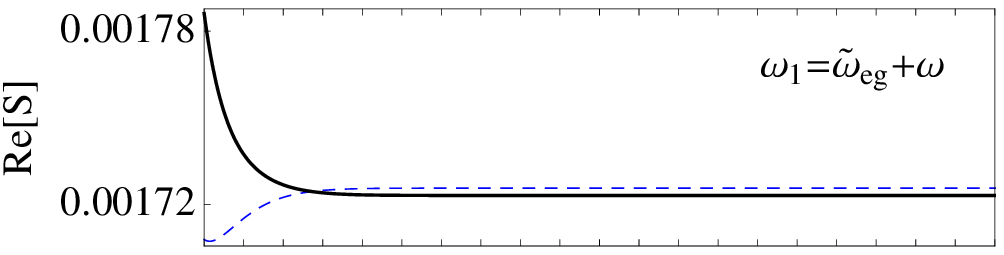}
\includegraphics[width=1\columnwidth]{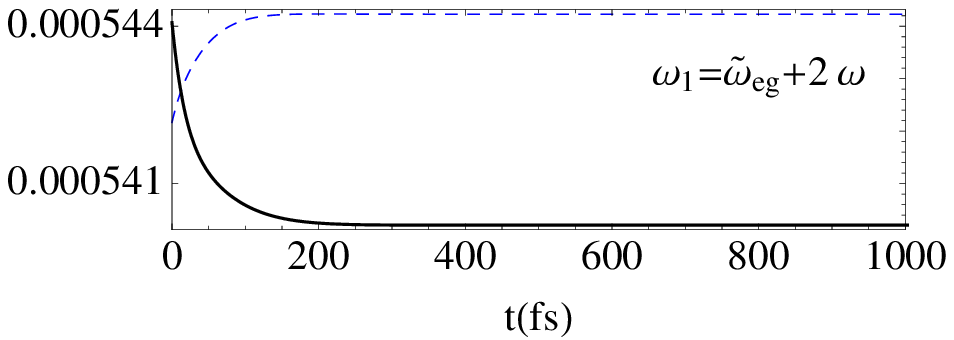}
\end{minipage}
\end{center}
\vspace{-.2cm}\begin{minipage}{0.45\columnwidth}
\includegraphics[width=.9\columnwidth]{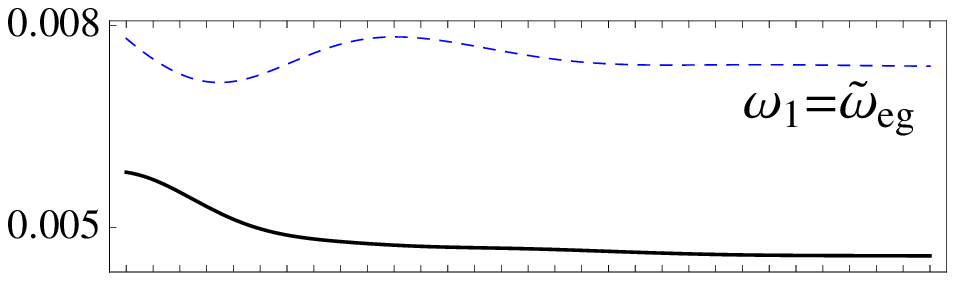}
\includegraphics[width=.9\columnwidth]{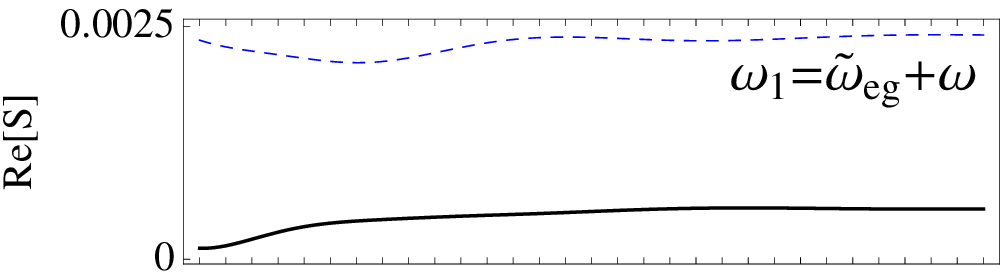}
\includegraphics[width=.9\columnwidth]{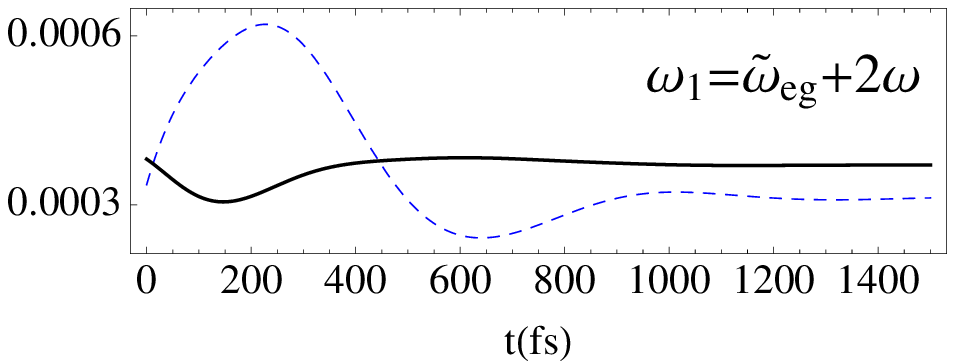}

\end{minipage}
\begin{minipage}{0.45\columnwidth}
\includegraphics[width=\columnwidth]{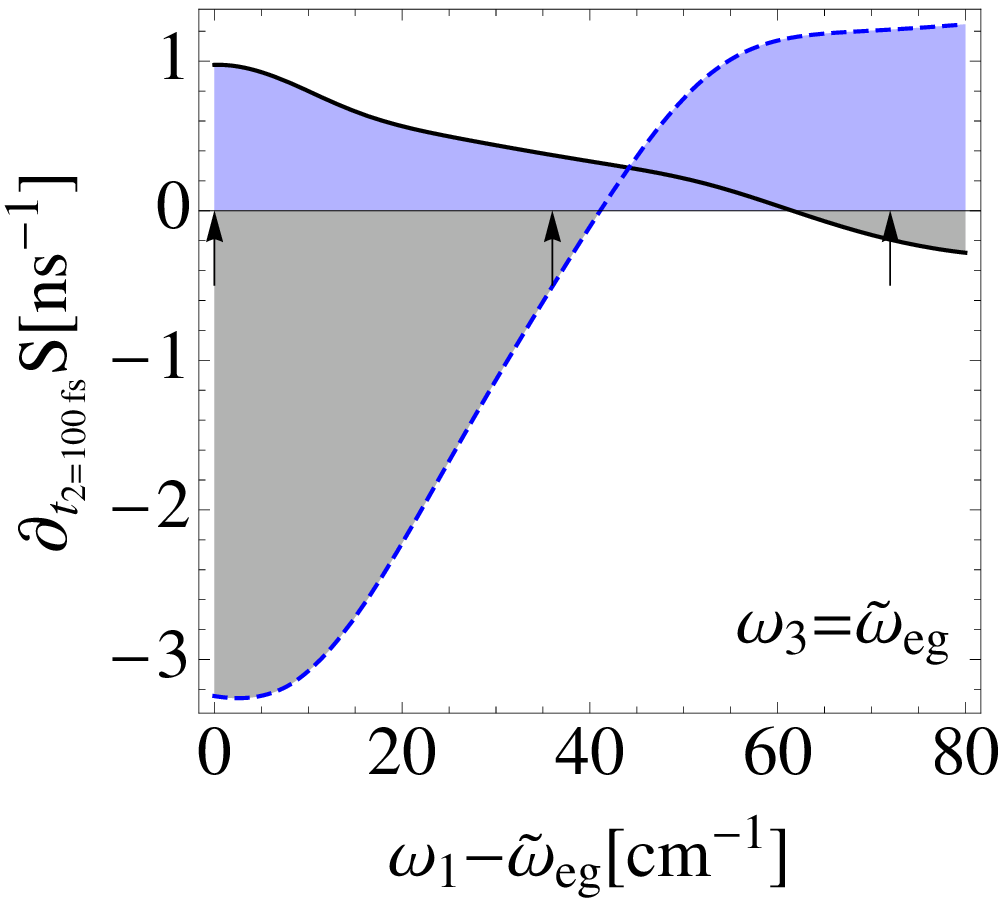}
\end{minipage}

\caption{In (a), Feynman diagrams involving the phonon-side band. Letters label electronic states, the superscript the boson state. In (b) are shown the waiting time domain signals at the homogeneously broadened peak at $\omega_1=\tilde{\omega}_{eg}+\omega$, $\omega_3=\tilde{\omega}_{eg}$ for the set $\{\gamma_0,\gamma_1,\gamma_2\}= \{6.2,8.1,11.2\}$ (left panel) and  $\{\gamma_0,\gamma_1\gamma_2\}= \{50,136,156\}$ cm $^{-1}$ (right panel). In (c) are presented the results for inhomogeneous broadened peaks  with $\sigma_E=102$ cm$^{-1}$, $\sigma_\omega=10$ cm$^{-1}$: in the left panel are shown the signals resolved in waiting time of the ZPL (top), side-band (middle) and second harmonic (bottom) peaks; in the right panel the slope $\partial_{t_2} \mbox{Re}[S]_{t_2=100{\mbox{\tiny{fs}}}} $ (long enough to avoid overlap with the second pulse $t_2\ge16 $fs, the equality fulfilled for the pulses duration achieved in \cite{Nemeth2010}) is presented; arrows highlight the positions of the ZPL, side band and overtone.  In all plots  non-rephasing (continuous, black) and rephasing (dashed, blue) path signals. 
}\label{fig3}
\end{figure}

 The result of the second order expansion of the homogeneously broadened 2D spectra for exciton 1 interacting with a vibrational mode at different waiting times $t_2$, is presented in Fig.\ref{fig1}(a)-(b) for the rephasing and non-rephasing contributions, respectively. A hyperbolic sine scale is used to highlight small features. First, note that in these figures, an asymmetry is prominent in the absorption dimension. This is highlighted in the zero-phonon emission profile (continuous line), where the difference in height of the peaks at  $\omega_3=\omega_{eg}$ and $\omega_1=\tilde\omega_{eg}\pm\omega$ accounts for  Stokes and anti-Stokes contributions due to  differences among populations of the initial canonical state in the vibrational manifolds, $\rho=\sum_n e^{-\beta n\omega}\ket{n}\bra{n}/\mbox{Tr}\{\cdot\}$.

Second,  notice that the rephasing and non-rephasing contributions to the response function are sensitive to different processes on the vibrational manifold at a given waiting time. In detail, at zero waiting time the vibrational structure around the ZPL peak is prominent in the rephasing spectra, where even the overtone is conspicuous  at $\omega_1=\tilde{\omega}_{eg}+2\omega$, while such structures are almost inexistent  in the non-rephasing signal at the same waiting time. 
For a waiting time $t_2\cong \pi/\omega$, corresponding to half a vibrational period, the behaviors are exchanged leading to a non-rephasing contribution with very well resolved side-bands \cite{Mancal2010}.  

 A better contrast  of the side-bands is possible in homogeneously broadened spectra, when instead of following the established procedure eq.(\ref{signal}) of adding the rephasing and non-rephasing responses (Fig.\ref{fig1}(c)), we proceed with their subtraction (Fig.\ref{fig1}(d)).The relative phases of rephasing and non-rephasing signals induce a stronger ZPL when added, but  strongly suppress the weight of the ZPL  to make the side-band structure gain relative weight when subtracted. This example allows to point out that the relative phases of the components that lead to the total signal are sensitive to algebraic operations over the individual contributions. However, this scenario  where the average among the ensemble is not performed, neglects the static inhomogeneity of importance in the ensemble required to individually address rephasing and non-rephasing contributions in different spatial directions.

In the following,  the exciton energy inhomogeneities are studied, considering the second order expansion of the line width function. The localized nature of  exciton 1 (at pigment 3) has been confirmed with numerical simulations   \cite{amerongenbook}, using a site energy distribution width of 110 cm$^{-1}$ which due small exchange narrowing leads to a value of $\sigma_E=102$ cm$^{-1}$ for the total width of the B825 band. Recent  numerical simulations of the FMO 2D spectra used a comparable  width  for the exciton 1 energy distribution (100 cm$^{-1}$)  \cite{Shi2011}.   The diagonal 1-1 peak for rephasing and non-rephasing signals is presented in Fig.\ref{fig2}(a)-(b), respectively, where the inhomogeneous broadening is introduced according to a gaussian distribution  of the excitonic energy with standard deviation $\sigma_E=102$ cm$^{-1}$. Interestingly, a very dramatic change occurs when moving from the homogeneous to the inhomogeneously broadened ensemble, where each path (rephasing/non-rephasing) is sensitive to addition of signals arising from different processes.  On the one hand, Fig.\ref{fig2}(a) shows the zero-phonon emission line (continuous) of the rephasing component, in order to highlight side-bands located at $\omega_{1}-\omega_3=\pm\omega$. On the other hand, the non-rephasing part, Fig.\ref{fig2}(b), is sensitive  to the overtone, highlighted by the continuous lines over the surface at the ZPL and overtones at $\omega_{1}-\omega_{3}=\pm2\omega$. A better contrast of the vibronic structure is accomplished in the non-rephasing 2D spectra due to the  drastic reduction of the ZPL. Hence, the  inclusion of inhomogeneous broadening does not simply regards to a general widening of the homogeneous resonances breath, since it affects in a different fashion to each of the resonances. In short, a discussion of this  reduction will allow us to understand the qualitative differences among homogeneous and inhomogeneously broadened spectra.

The inhomogeneous broadening in the excitonic frequencies corresponds to widening along the diagonal in the 2D spectra, with a ZPL maximum  at $\omega_1=\omega_3$. We have chosen  the waiting time long enough in Fig.\ref{fig2}(a)-(b), to avoid the $t_2$ dependence on the 2D spectra. These results remain valid for different waiting times as will be demonstrated below. The  expression for the zeroth order (waiting time independent) contribution of the ZPL reads:
\begin{widetext}
\begin{eqnarray}
 S^0(\omega_1&=&\omega_3)_{\mbox{\tiny{ZPL}}}=\frac{1}{(\gamma_0^2+(\omega_1-E_{\psi_1})^2)^2} ( \gamma_0^2\pm(E_{\psi_1}-\omega_1)^2+i\gamma_0(\omega_1-E_{\psi_1}\mp \omega_1\pm E_{\psi_1})),
\end{eqnarray}
\end{widetext}
where the upper and lower signs correspond to rephasing and non-rephasing expressions, respectively. The real part of the rephasing contribution is always positive for any value of the exciton 1 energy $E_{\psi_1}$, while that of the non-rephasing part,  is positive for $E_{\psi_1}-\omega_1<\gamma_0$   and negative otherwise. The imaginary part of the rephasing signal is zero at the diagonal, while that of the non-rephasing, is positive for $E_{\psi_1}<\omega_1$ and negative otherwise. This situation leads to have at a single point in the diagonal of the 2D spectra, only positive contributions for the rephasing part  that will add up to build the ZPL peak. On the other hand, in the non-rephasing path the signals have varying signs depending on the exciton energy, which when added up  in the inhomogeneous sample, will lead to a decreased ZPL. This fact is quantified in Fig.\ref{fig2}(c)-(e), where the weighted contribution $e^{-(E_{\psi_1}-\tilde{\omega}_{eg})^2/2\sigma_E^2}S(\omega_1,\omega_3,E_{\psi_1})$ of the complete second order expansion as a function of the difference $E_{\psi_1}-\tilde{\omega}_{eg}$ are presented. The Fig.\ref{fig2}(c)  shows that the signal at $\omega_1$=$\omega_3$=$\tilde{\omega}_{eg}$ has alternating (equal) signs for real and imaginary parts of the non-rephasing (rephasing) signal.  The trend continues (Fig.\ref{fig2}(d)) for points regarding the phonon side-band $(\omega_1=\tilde{\omega}_{eg}+\omega,\omega_3=\tilde{\omega}_{eg})$ where the real (imaginary) part of the rephasing signal presents  over a great range of the whole inhomogeneous distribution, positive (negative) values that constructively add to build up a appreciable side band in the rephasing path.  However, rephasing and non-rephasing parts have positive and negative contributions along the inhomogeneous distribution from the signals arising at the overtone  resonance $(\omega_1=\tilde{\omega}_{eg}+2\omega,\omega_3=\tilde{\omega}_{eg})$ which lead to a reduction commensurate in both.  The concomitant reduction of the side band and drastic reduction of the ZPL for the non-rephasing direction, explains the surprising enhancement in the contrast for the signal arising from overtone, due to the (usually  troublesome)  sample inhomogeneities.

The inhomogeneities are not only ascribed to the exciton energy, but also the vibrational manifold.  For instance, hole burning can select a subensemble having an optical transition frequency, but the inhomogeneities in the satellite structure due to the vibrational manifold will not be different from those of any other subensemble. Therefore, the width of the satellite structure in these experiments reflects that of the ensemble.  The low energy satellite structure in the B825 was observed in hole burning \cite{Small2000},  with a width of 10 cm$^{-1}$ for the $\omega=36$ cm$^{-1}$ resonance.  Hence, we proceed to perform an average of the vibrational frequency, according to a gaussian with $\sigma_\omega=10$ cm$^{-1}$, and present the results for rephasing and non-rephasing paths in Fig.\ref{fig2}(f)-(g), respectively. In the rephasing contribution, this average smears out the vibronic structure, while some of it remains in the non-rephasing  path.  Since the experiment involve macroscopic samples, the  vibrational manifold ensemble inhomogeneities will lead to a reduced structure in the 2D spectra, and a simple inspection will not permit (in general) a straightforward inference of the presence of discrete vibronic contributions.
 Furthermore, the sampling  of the signal in $t_1$ is done during a finite time that imposes a minimum frequency, hence a maximum resolution among frequencies. Currently, studies of 2D spectra tracking the vibrational wave-packet motion, used a scanning range for $t_1$ from -100 to 100 fs, which allows a resolution in $\Delta\omega_1\simeq 52$ cm$^{-1}$ for rephasing/non-rephasing signals, or half of it,  $\Delta\omega_1\simeq 26$cm$^{-1}$ for the total signal \cite{Nemeth2010}.
Accordingly, the discovery of vibronic structure in the ensemble becomes difficult relying on the inspection of a single 2D spectrum alone. However, the vibronic contributions can be detected from  the waiting time $t_2$, which induces features that allow us to infer  the presence of discrete phonon modes.

The additional structure whose peaks lie at $\omega_1-\omega_3=\pm n \omega$, imply rotations occurring in the vibrational manifold that, as will be seen,  induce a phase shift in the signal on waiting time dimension. To illustrate our point, we start with the second order line-width expansion at the homogeneous level and focus on the real part of the signal concerning the product of absorptive (non-dispersive) spectra in $\omega_1$ and $\omega_3$. 

For instance, at $\omega_1=\tilde{\omega}_{eg}+\omega$,   $\omega_3=\tilde{\omega}_{eg}$ the signal is constructed from the Liouville paths contribution  shown in the Feynman diagrams of Fig.\ref{fig3}(a), where time evolves from bottom to top over the vertical lines that represent the ket and bra evolution. The wiggly arrows denote interactions with electromagnetic pulses which excite (de-excite) the system when they are directed into (out of) the ket-bra vertical lines. The additional kets or bras, are used to clarify the transitions involved after a given excitation.  The diagrams $R_1$ and $R_2$ ($R_3$ and $R_4$) involve rotations in the excited (ground) state manifold for the waiting time domain. Observation of the point $\omega_1=\tilde{\omega}_{eg}+\omega$ is generated when the bra (ket)  $\bra{e^{n+1}}$ ($\ket{e^{n+1}}$) is excited with the first pulse. On the other hand, at $\omega_3=\tilde{\omega}_{eg}$ the evolution involves the same boson quanta in both bra and ket. Therefore a flip of the boson state $n+1\rightarrow n$ is required in order to end up in a population ket-bra combination before the trace operation is performed. For diagrams $R_3$ and $R_4$ this rotation can be accomplished when the second pulse de-excites the system to $\bra{g^{n+1}}$  and $\ket{g^{n+1}}$, respectively. However, for $R_1$ and $R_2$ the rotation must happen in the waiting time, and therefore, the signal is sensitive to the the required flip in the vibrational manifold. This rotation can be explained along the mentioned mathematical procedure, but can also be  understood by considering the Frank-Condon principle. 

After  pulsed excitation,   the system is left in a superposition of eigenstates of the Hamiltonian (\ref{Heph}),  which also represent a  vibronic progression  $\ket{\tilde{n}}$ whose energy differ by $\omega$ among consecutive levels  in a given electronic manifold. As the thermal population involves the lowest energy levels, the contribution to the signal will primarily arise from  the ZPL and the phonon side-band, approximately described by  $\ket{\phi}\approx \alpha_0 \sin(\omega t)\ket{\tilde{0}}+ \alpha_1\cos(\omega t)\ket{\tilde{1}}$.  Simply stated,  the signal at $\omega_{1(3)}=\tilde{\omega}_{eg}\pm n\omega$   is mainly contributed by the $\ket{\tilde{n}}$ bosons state population, as a witness of excitation and dynamics of the vibrational wave-packet motion. The phase of the signal will vary as the 2D spectrum is swept  reflecting the phase difference among populations of the levels being scanned.  In the waiting time, Fig.\ref{fig3}(b), the signal presents the discussed  phase shift $\approx \pi$ among rephasing and non-rephasing contributions (remind Fig.\ref{fig2}(a)-(b)) . However, this panel also shows a phase shift among ZPL and side band signals of $\pi$, to resemble the just mentioned population inversion. A population picture can be misleading for the phase difference among the side band and overtone, since this latter resonance develops a signal with strength proportional to $s^2$, small enough to be contaminated by the tails of the other resonances. The right panel  of Fig.\ref{fig3}(b), shows that if the breath of the homogeneous broadened resonances  is amplified in accordance with loosing the mode discreteness, the phase shift among different points in the 2D spectrum is lost when scanning the neighborhood of a diagonal peak. Hence, a smeared out interference of several indistinguishable contributions in a given area of the 2D spectrum reduces the phase shift differences. 

In Fig.\ref{fig3}(c) left panel,  is shown the result of averaging static inhomogeneity to the signals resolved in the $t_2$ dimension. Here it can be first noticed the reduction of the ZPL and side band of the non-rephasing compared to rephasing path. Secondly, this procedure only slightly reduces the phase difference among rephasing and non-rephasing components. Third,  the phase differences among resonances on a given path is persistent among ZPL, side band and overtone.  A variation in the phase leads also to a variation of the slope $\partial_{t_2}\mbox{Re}[S](\omega_1,\omega_3)$ at a fixed $t_2$, before the settlement of the stationary state. This variation is presented in the right panel of Fig.\ref{fig3}(c), and specially highlights (by filling the area underneath) the change of sign of the signal slope  for a fixed waiting time as a function of the coordinate $\omega_1$. Even though the actual values of this phase are affected by the presence of the ensemble inhomogeneities, its variation remains, and since its origin is traced back to the  discreteness of the vibrational manifold, it can be used as a tool to diagnose the presence of vibronic discrete structure beyond the unavoidable presence of static disorder. 

The 2D spectra consist of diagonal and non-diagonal peaks. We have covered the behavior of signatures in both homogeneous and inhomogeneously broadened samples for the diagonal peaks. Next, we consider a prominent  signature that arises in the waiting time evolution of non-diagonal excitonic peaks.  \\The single excitation peak on the model just presented is unable to capture the behavior of excitonic coherence. Moreover, it has been previously stated  \cite{FMOnature,Engel2011}, that extracting the dynamics of non-diagonal peaks in non-linear spectra, that is the excitonic coherences, has the advantage of less congested spectra. Therefore, it is an important contribution to understand the features that will arise in the  cross peak dynamics on the FMO light harvesting complex.

\begin{figure*} 
\includegraphics[width=.49\columnwidth]{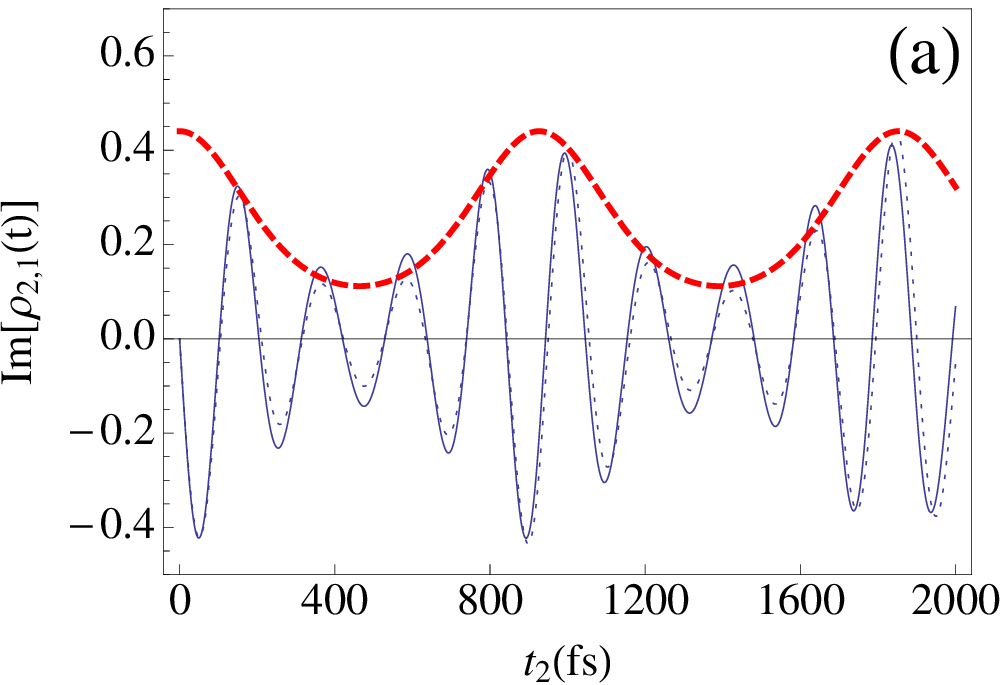}\hspace{.8 cm}
\begin{minipage}{.4\columnwidth}

\setlength{\unitlength}{0.7mm}
\vspace{-2.5 cm}\begin{picture}(50,45)
\put(-10,0){\line(0,1){47}}
\put(50,0){\line(0,1){47}}
\put(-10,0){\line(1,0){60}}
\put(-10,47){\line(1,0){60}}
\put(-9,43){(b)}

\put(2.6,6){$_{\ket{g}\bra{g}}$}
\put(5,8){\line(0,1){30}}
\put(11,8){\line(0,1){30}}
\put(5,21){\ldots}
\put(5,15){\ldots}
\put(7,18){$_{t_1}$}
\put(7,24){$_{t_2}$}
\put(7,30){$_{t_3}$}
\put(5,27){\ldots}
\put(5,33){\ldots}
\put(2,18){\begin{rotate}{30}$\leadsto$\end{rotate}}
\put(1.7,25){\begin{rotate}{30}$\leadsto$\end{rotate}}
\put(15.4,12){\begin{rotate}{150}$\leadsto$\end{rotate}}
\put(5.5,34){\begin{rotate}{150}$\leadsto$\end{rotate}}
\put(1.8,41){$_{\ket{\psi_1}\bra{\psi_1}}$}
\put(-2,23){$_{\ket{\psi_2}}$}
\put(12,17.5){$_{\bra{\psi_1}}$}
\put(-1.5,218.5){$ _{k_2}$}
\put(12,9.5){$ _{-k_1}$}
\put(-2,28){$_{k_3}$}
\put(-2.4,37.5){$_{-k_s}$}
\put(5,-0.5){$$}

\put(27,6){$_{\ket{g}\bra{g}}$}
\put(30,8){\line(0,1){30}}
\put(36,8){\line(0,1){30}}
\put(30,21){\ldots}
\put(30,15){\ldots}
\put(32,18){$_{t_1}$}
\put(32,24){$_{t_2}$}
\put(32,30){$_{t_3}$}
\put(30,27){\ldots}
\put(30,33){\ldots}
\put(27,18){\begin{rotate}{30}$\leadsto$\end{rotate}}
\put(36.7,26){\begin{rotate}{30}$\leadsto$\end{rotate}}
\put(40.4,12){\begin{rotate}{150}$\leadsto$\end{rotate}}
\put(30,35){\begin{rotate}{150}$\leadsto$\end{rotate}}
\put(27,41){$_{\ket{g}\bra{g}}$}
\put(23,28){$_{\ket{\psi_2}}$}
\put(37,17.5){$_{\bra{\psi_1}}$}
\put(23.5,18.5){$ _{k_2}$}
\put(37,9.5){$ _{-k_1}$}
\put(39,31){$_{k_3}$}
\put(23,37.5){$_{-k_s}$}
\put(30,-.5){$$}
\end{picture}
\end{minipage}
\includegraphics[width=.6\columnwidth]{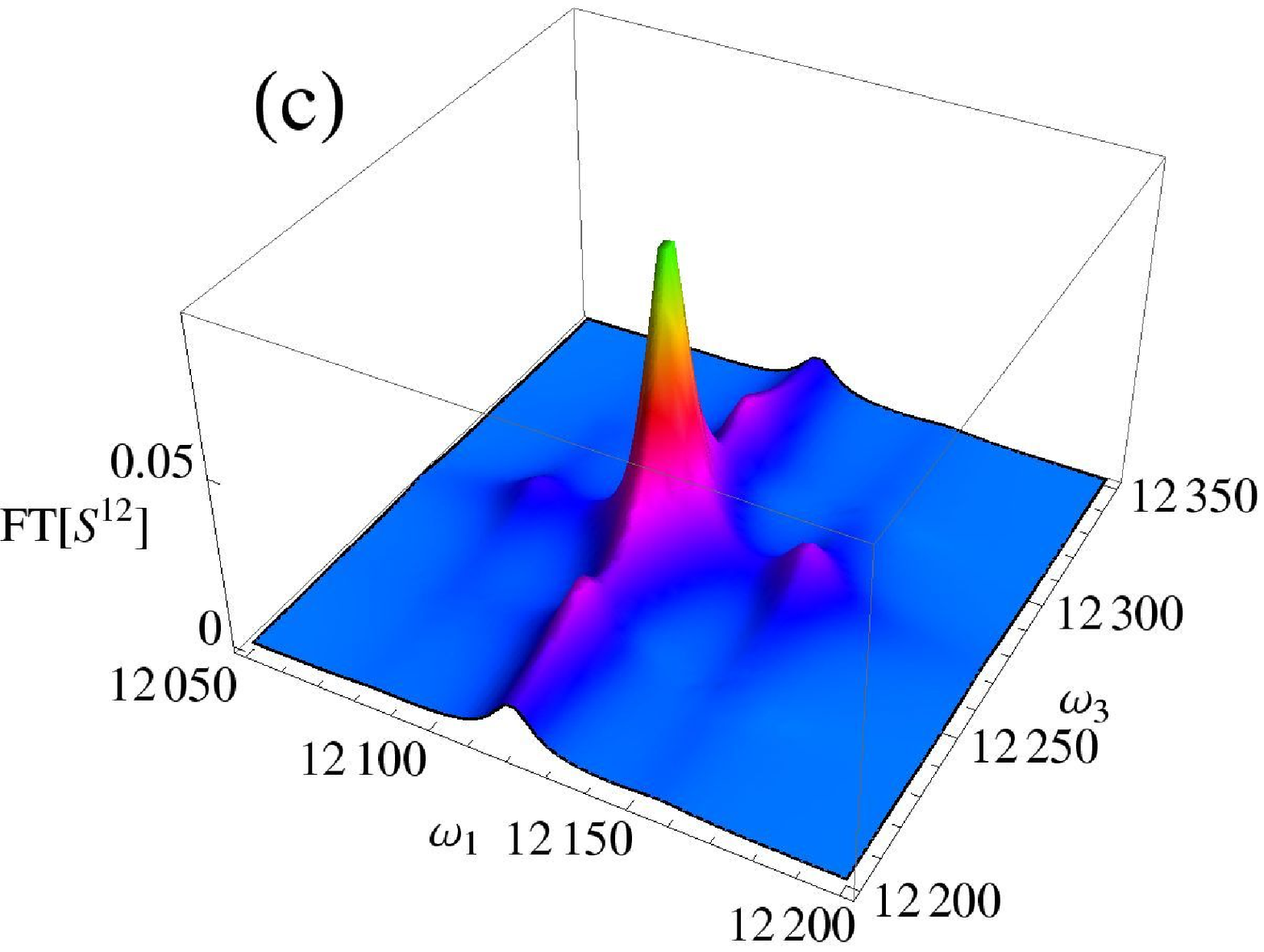}\vspace{.4 cm}
\includegraphics[width=.6\columnwidth]{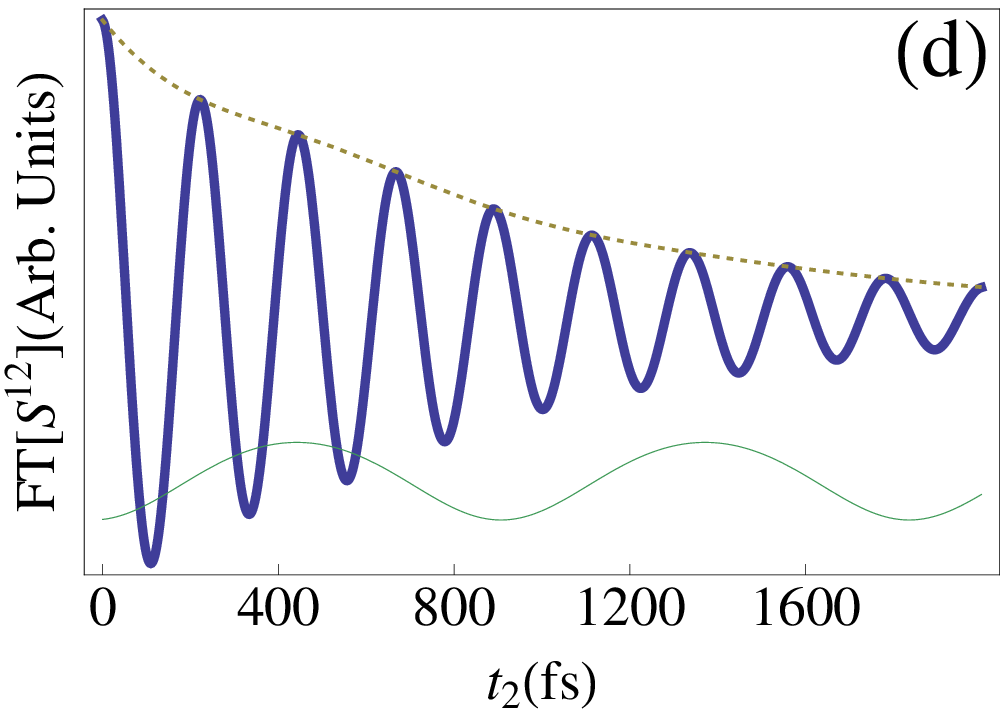}
\includegraphics[width=.6\columnwidth]{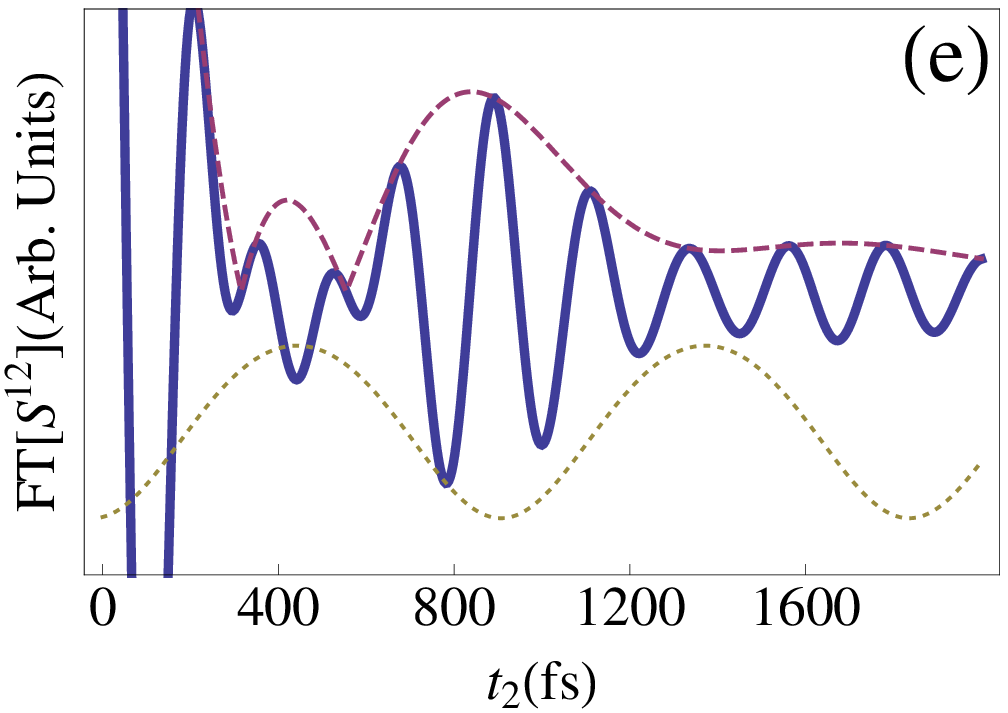}
\caption{Excitonic coherence $\rho_{2,1}(t)=\mathrm{Tr}\{\ket{\psi_1}\bra{\psi_2}\rho(t)\}$. In (a) full numerical calculation (continuous) and electronic coherence analytical result (dotted). In (b) are shown the Feynman diagrams that contribute to oscillating terms at frequencies concerning the excited electronic states on the response function. In (c) the result of the homogeneously broadened signal is presented for the the addition of both paths,  at a waiting time $t_2=20$ fs. In (d)-(e), the waiting time resolved signal at points $\{\omega_1,\omega_2\}=\{\tilde{\omega}_{\psi_1},\tilde{\omega}_{\psi_2}\}$ and $\{\tilde{\omega}_{\psi_1}+\omega,\tilde{\omega}_{\psi_2}\}$ respectively, is presented. Thick continuous line represent the real part of these oscillatory contributions.  For comparison purposes, we show the absolute value of the rephasing signal (dotted) and the amplitude envelope from electronic coherence solution eq.(\ref{coherence}) (thin, continuous) scaled in amplitude and shifted in phase as described in the text.}\label{fig4}
\end{figure*}

\subsubsection{ Vibronic features in 2D spectra: Non-diagonal peaks} In the following, we investigate the excitonic coherence of a dimer interacting with localized vibrations in order to address  the FMO exciton 1-2 cross peak, when the coupling to the $\omega=36$ cm$^{-1}$ mode is accounted for.  The sites  with the greatest contribution on excitons 1 and 2, are chromophores historically \cite{Matthews1980}  named as sites 3 and 4\cite{Matthews1980}  (states $\ket{3}$ and $\ket{4}$). The Hamiltonian restricted to a single excitation for these two sites, each interacting with intramolecular vibrations is:
\begin{widetext}
\bea
H&=&H_{b1}+H_{b2}+\sum_{i=3,4}\omega_i\sqrt{s_i}X_i\ket{i}\bra{i}+ \sum_{i=3,4} \omega_{gi}\ket{i}\bra{i}+V(\ket{3}\bra{4}+\ket{4}\bra{3})\label{Hdimer}
\label{hlow}
\eea
\end{widetext}
where the last two terms correspond to the electronic Hamiltonian, and $H_{bi}=\omega_{i}b_{i}^{\dagger}b_{i}$,  $X_{i}=b_{i}+b_{i}^{\dagger}$, $b_i$ ($b_i^+$) are annihilation (creation) operators of the ith vibration. Diagonalization of the electronic Hamiltonian leads to excitonic states $\ket{\psi_1}=\frac{1}{\sqrt{1+a^2}}(\ket{3}-a\ket{4})$,  $\ket{\psi_2}=\frac{1}{\sqrt{1+a^2}}(a\ket{3}-\ket{4})$, parametrized by $a=(\sqrt{\delta^2+4V^2}-\delta)/2V$ which for $\delta= \omega_{g4}-\omega_{g3}$ describes the limiting cases of localized ($\vert V\vert \ll \delta$, $a\approx 0$) and fully delocalized ($\vert V\vert\gg \delta$,  $a\approx \pm1$) excitons.\\
In order to determine the effect of the vibrational modes in the exciton dynamics in the particular case of the FMO excitons 1 and 2, we proceed with a numerical solution constrained to the experimentally determined parameters. The dipole-dipole couplings in agreement with linear absorption, circular dichroism and linear dichroism  \cite{Rengereview}, and the experimentally determined exciton transition energies \cite{Engel2011} of the full FMO, were used in an iterative procedure as described in Ref.\cite{Engel2011} to determine the site energies of $12142$ cm$^{-1}$ and 12315 cm$^{-1}$ for sites 3 and 4, respectively. Then, the value $a=0.28$ is determined from the dipole-dipole coupling (-$53$ cm$^{-1}$, \cite{Rengereview}) between sites 3 and 4. We constrain the reduced dimer model to match the delocalization degree $a=0.28$ and the experimentally determined exciton 1 and 2 energy difference  $\delta_{\psi_1\psi_2}=158$ cm$^{-1}$ \cite{Engel2011}, which results in  renormalized site energies  $\delta=135$ cm$^{-1}$ and coupling strength $V=-41$cm$^{-1}$.  Given that our interest is the  1-2 cross peak of FMO whose beating signal is a manifestation of excitonic coherences, Fig.\ref{fig4}(a) shows the result of the numerical calculation of excitonic coherence. In the absence of electron-vibration coupling, the amplitude of the coherence should be monotonically decreasing, since a dimer model will only have a single excitonic frequency in the one  excitation manifold. The most prominent manifestation of the presence of the vibronic structure is observed here, as a modulation of the excitonic coherence amplitude. In order to provide  a figure of merit beyond a qualitative estimation for this modulation, an analytical study is much preferable.\\
In the excitonic basis, the projector $\ket{3}\bra{3}=\frac{1}{1+a^2}(\ket{\psi_1}\bra{\psi_1}+a^2\ket{\psi_2}\bra{\psi_2}+a\ket{\psi_1}\bra{\psi_2}+a\ket{\psi_2}\bra{\psi_1})$ with a similar expression for $\ket{4}\bra{4}$ by the exchanges $\psi_1\rightarrow-\psi_2$ and $\psi_2\rightarrow\psi_1$, shows that in situations where $\delta$  is  much greater than the  Coulomb interaction, the Hamiltonian eq.(\ref{hlow}) is diagonal in the excitonic basis.  With this approximation the excitonic coherence $\mathrm{Tr}\{\ket{\psi_1}\bra{\psi_2}\rho(t)\}$ can be calculated using the total density matrix evolution $\rho(t)=e^{iHt}\rho(0)e^{-iHt}$. Given that our interest lies in the dynamics on the waiting time domain, we suppose an initial state $\rho(0)$ which is  a product of  the $i$th oscillator  thermal state $\rho_{bi}$ and an electronic  superposition  proportional to the site transition dipole moments, $\rho^{e}\simeq \sum_{i,j}\mu_i\mu_j\ket{i}\bra{j}=\sum_i\mu_{\psi_i}\mu_{\psi_j}\ket{\psi_i}\bra{\psi_j}$ to best describe the situation after the arrival of two broad band  low intensity pulses within the single excitation subspace. After tracing out  the exciton manifold and then noting that the trace over the oscillators can be arranged as a product of traces due to commutation among operators describing vibrations in different chromophores,  we make use  of the displacement operator $U_{i}=\exp[-s_i(b_{i}-b_{i}^{\dagger})]$ property:  $U_{i}\,(H_{bi}+\sqrt{s_i}\omega_iX_{i})U_{i}^{-1}=H_{b1}-s_i\omega_{i}$. This procedure  leads to the interaction picture displacement operator  $U_{i}(t)=e^{iH_{bi}t}U_{i}e^{-iH_{bi}t}$ correlation function  $\mathrm{Tr}_{bi}\left\{U_{i}^{-1}(-t)U_{i}\rho_{bi}\right\}$, which for harmonic oscillators, is a standard result \cite{Mukamel}. Hence follows:
\begin{align}\label{coherence}
&\mathrm{Tr}\left\{\ket{\psi_1}\bra{\psi_2} \rho(t)\right\}=\frac{\mu_{\psi_1}\mu_{\psi_2}}{\mu_{\psi_1}^2+\mu_{\psi_2}^2}e^{-i\delta_{\psi_1\psi_2}t+g_3(t)+g_4^*(t)}\nonumber\\
&=\frac{\mu_{\psi_1}\mu_{\psi_2}}{\mu_{\psi_1}^2+\mu_{\psi_2}^2}e^{-i\delta_{\psi_1\psi_2}t-2s\coth(\beta\omega/2)(1-\cos(\omega t))}
\end{align}
where $g_i(t)$ are the linewidth functions (remind eq.(\ref{linewidth})) addressing the properties of vibrations in the ith site. As shown in the second line of the above equation, if oscillators are identical, the addition of the linewidth functions  exp$[g_3(t)+g_4^*(t)]$ results in a real argument exp$[-2s\coth(\beta\hbar\omega/2)(1-\cos(\omega t))]$ presented (dashed)  in Fig.\ref{fig4}(a), that quantitatively reflect the amplitude modulation of the excitonic coherence.  Also the complete solution, eq.(\ref{coherence}) in dotted lines shows excellent agreement with full numerical simulations. This underline the fact that disregarding excitonic population transfer terms mediated by vibrational modes $\propto \frac{a}{1+a^2}\ket{\psi_i}\bra{\psi_j}$ seems correct in the B825, for the description of the vibronic wave packet motion. Moreover, this  agreement in the current  situation where $V\simeq\delta$, highlights that in a more general context,  the periodic modulation of the excitonic coherence, i.e., of the signal amplitude, is a signature of  coherent dynamics occurring in the quantized vibrational manifold. The analytical result shows that the modulation of amplitude in the excitonic coherence will present a frequency equal to that of the prominent vibrational mode, namely $\omega$, and will have a depth  proportional to the coupling of such vibration with the electronic transitions, via the Huang-Rhys factor $s$.  \\
Therefore, for practical purposes, the assumption that the Hamiltonian eq.(\ref{Hdimer}) is diagonal in the exciton basis when accounting on vibronic effects on the 1 and 2 excitonic coherence in the FMO, is well supported. We proceed with this assumption to investigate whether the modulation of excitonic coherence amplitude can be readily observed in the 1-2 FMO cross-peak from 2D spectra. The diagrams in Fig.\ref{fig4}(b)  are the contributions for the excitonic cross peak that present excited state evolution in the waiting time at a frequency equal  to the energy difference $\tilde{\omega}_{\psi_2}-\tilde{\omega}_{\psi_1}=\delta_{\psi_1\psi_2}$, with corrections arising from the line-width functions. These oscillating terms  carry dynamical information that provide access to the interference of the most salient dynamical contributions, and read:
\begin{align}\label{eqrev}
S^{12}(&t_1,t_2,t_3)=2\,\mbox{Im}[\exp[-i\tilde{\omega}_{\psi_1}t_1+i\tilde{\omega}_{\psi_2}t_3+i \delta_{\psi_1\psi_2}t_2]\nonumber\\
&\times ( \mu_{\psi_1}\mu_{\psi_2}\mu_{\psi_2,f}\mu_{f,\psi_2}\exp[f_1(t_1,t_2,t_3)] \nonumber\\
& \,\,\,\,\,\,\,\,\,+ \mu_{\psi_1}^2\mu_{\psi_2}^2 \exp[f_2(t_1,t_2,t_3)])]
\end{align}
where $\mu_{\psi_i,f}$ is the dipole moment of the transition $\ket{\psi_i}\rightarrow\ket{f}$, and $f_{1(2)}(t_1,t_2,t_3)$ correspond to functions associated with the cumulant expansion of the linewidth functiond of the paths shown in Fig.\ref{fig4}(b) under the assumption of a diagonal exciton-phonon interaction \cite{Cho}:
\begin{widetext}
\begin{eqnarray}
f_1(t_1,t_2,t_3)&=&-g_{\psi_2\psi_2}^*(t_2)-g_{ff}^*(t_3)-g_{f\psi_1}(t_1+t_2)+g_{\psi_2f}^*(t_2)+g_{\psi_2f}^*(t_3)+g_{\psi_2\psi_1}(t_1+t_2)-g_{\psi_2\psi_1}(t_1)-g_{\psi_2\psi_1}^*(t_3)\nonumber\\
& &+g_{f\psi_1}(t_1+t_2+t_3)-g_{\psi_1\psi_1}(t_1+t_2+t_3)+g_{\psi_2\psi_1}^*(t_2+t_3)-g_{\psi_2f}^*(t_2+t_3)+g_{f\psi_1}^*(t_3),\nonumber\\
\vspace{1.6 cm}f_2(t_1,t_2,t_3)&=&-g_{\psi_1\psi_1}^*(t_1+t_2)-g_{\psi_1\psi_2}^*(t_1)+g_{\psi_1\psi_2}(t_2)+g^*_{\psi_1\psi_2}(t_1+t_2+t_3)-g_{\psi_1\psi_2}^*(t_3)-g_{\psi_2\psi_2}(t_2+t_3)\nonumber
\end{eqnarray}
\end{widetext}  
These expressions involve photon-echo rephasing contributions with linewidth functions of the singly excited exciton states $\ket{\psi_1},\ \ket{\psi_2}$ and doubly excited states $\ket{e_3,e_4}=\ket{f}$. The transformation of the coordinates $X_i$  in the exciton basis yields to $X_f=X_3+X_4, X_{\psi_1}=\frac{1}{1+a^2}(X_3+a^2 X_4), X_{\psi_2}=\frac{1}{1+a^2}(a^2 X_3+ X_4)$, which allows calculation of  the linewidth function $g_{ij}(t)$ in terms of the site linewidth function $g(t)$. For identical vibrations in each site, it can be written $g_{ij}=C_{ij}g(t)$, with coefficients $\{C_{ff},C_{\psi_i f}=C_{f\psi_i},C_{\psi_i \psi_j},C_{\psi_i \psi_i}\}=\{2,1,\frac{2a^2}{(1+a^2)^2},\frac{1+a^4}{(1+a^2)^2}\}$ for $i\ne j$. When $a\rightarrow0$, the above expressions fulfill $f_1(t_1,t_2t_3)=f_2(t_1,t_2t_3)$, while $\mu_{\psi_1,f}=\mu_{\psi_2}=\mu_{4}$ and $\mu_{\psi_2,f}=\mu_{\psi_1}=\mu_{3} $ to produce the expected vanishment of non-diagonal peaks in absence of electronic coupling among sites.\\
Following the same procedure used for the diagonal peak in the homogeneously broadened  FT signal, we perform a  first order Taylor expansion of $e^{f_i(t)}$ and present in Figs.\ref{fig4}(c)-(e)  the result of the 2D  non-diagonal peak following eq.(\ref{eqrev}). The 2D spectra at early waiting times in Fig.\ref{fig4}(c), shows appreciable contribution from the side-band structure. At the coordinate $\{\omega_1,\omega_3\}=\{\tilde{\omega}_{\psi_1g},\tilde{\omega}_{\psi_2g}\}$ (Fig.\ref{fig4}(d))  the modulation of the signal envelope is absent and supports that its principal contribution will mainly arise from ZPL transitions.  On the other hand,  the signal arising from the satellite structure $\{\omega_1,\omega_3\}=\{\tilde{\omega}_{\psi_1}+\omega,\tilde{\omega}_{\psi_2}\}$ shows an appreciable modulation of its amplitude. This modulation has a frequency $\simeq\omega$, that can be accounted for in virtue of eq.(\ref{coherence}) to the vibrational wave packet motion.

The modulation in the signal of the cross peaks makes it a suitable witness for discrete vibronic structure in the environment. We follow with  the detailed study of the progression from homogeneous to inhomogeneous broadening of the signal,  which leads to interesting conclusions.    


\subsection{ Ensemble inhomogeneities: On the static nature of the difference among site and excitonic coherence decay} 
\begin{figure*}

\includegraphics[width=.7\columnwidth]{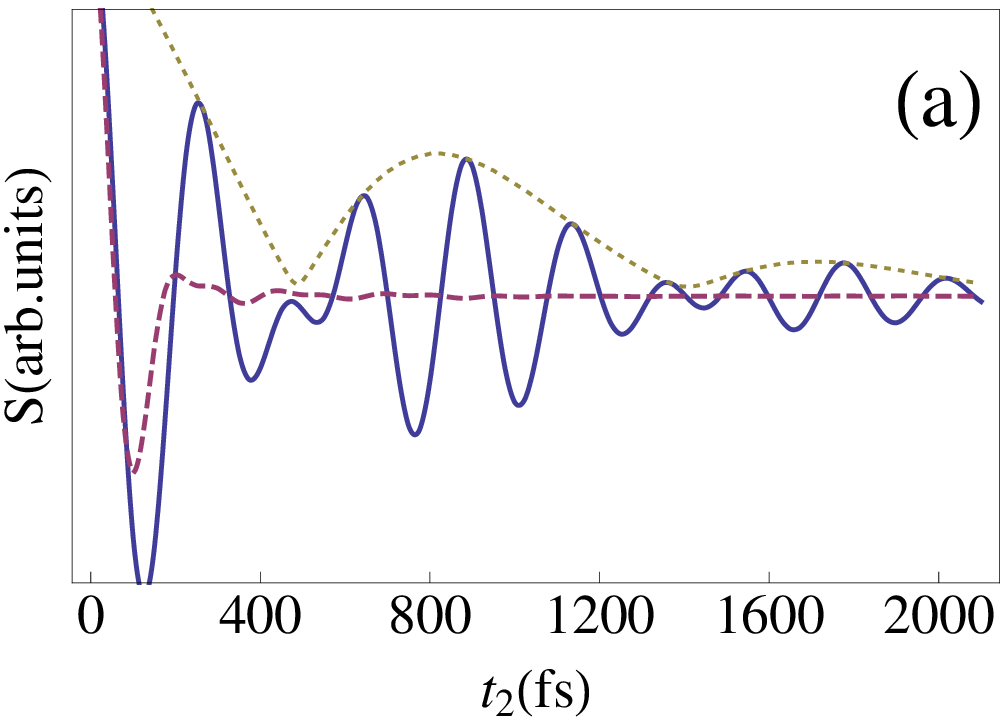}
\includegraphics[width=.7\columnwidth]{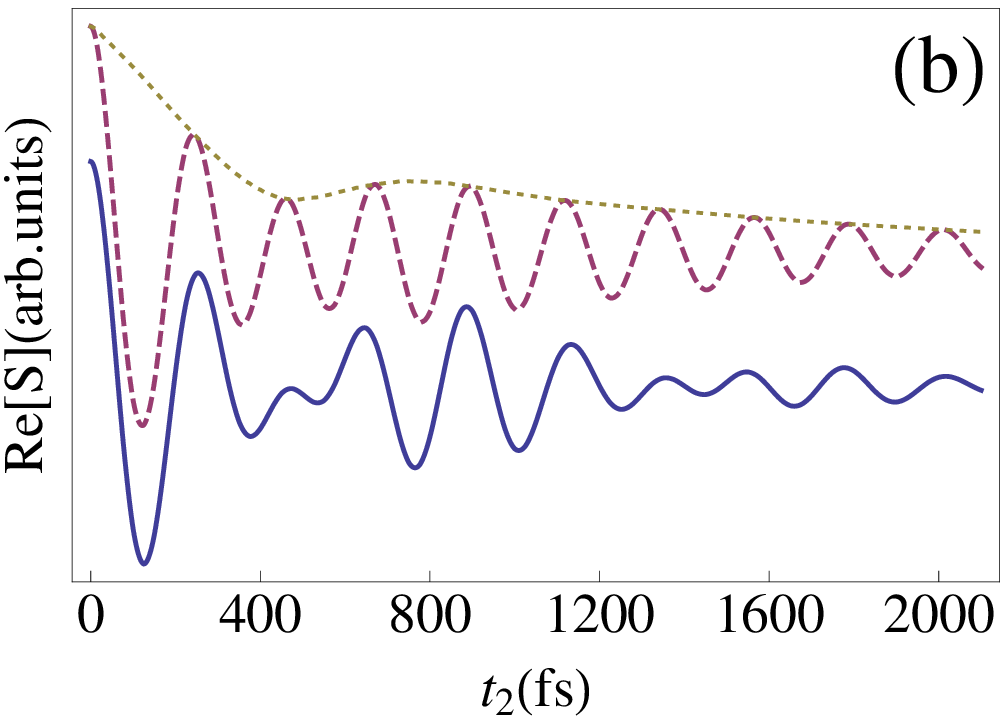}
\caption{ In (a), the result of averaging homogeneous contributions with independent energy variations equal to those reported \cite{Louwe1997} of $\{\sigma_{\epsilon},\sigma_{\epsilon^\prime}\}=\{102,80\}$ cm$^{-1}$ (dashed) and with excitonic energy variation with $\{\sigma_{\epsilon},\sigma_{\epsilon-\epsilon^\prime}\}=\{102,5\}$ cm$^{-1}$ (real part, continuous; absolute value dotted). In (b) the additional inhomogeneity arising from the vibronic ensemble is included, and the result of the averaging using  gaussian distributions with standard deviations $\sigma_\omega=\{1,10 \}$cm$^{-1}$ are presented in continuous and dashed lines, respectively. Dotted line is the absolute value for  $\sigma_\omega=10 $cm$^{-1}$. In all plots $s=0.24$ (see text).}\label{fig5}
\end{figure*}

Two different scenarios are shown in Fig.\ref{fig5}(a) for the introduction of static disorder. In the first case (dashed line), the  signal is averaged assuming uncorrelated static fluctuations, distributed according to the product of probability density functions from  individual energies $P(\tilde{\omega}_1)P(\tilde{\omega}_2)$, using comparable inhomogeneous widths on each. In such a case, the damping of the oscillations occurs in a time-scale $\approx 100-200$ fs, which  does not represent the experimentally measured result \cite{Engel2011b,EngelNJP}.  By virtue of the microscopic model proposed here, the homogeneous decay presented in Fig.\ref{fig4}(d)-(e) is  similar, while the decay  due to uncorrelated variation in the inhomogeneous average in Fig.\ref{fig5}(a) dashed line, is far too fast, compared to the appreciable visibility experimentally found for oscillations well beyond $t_2=1$ ps  \cite{Engel2011b,EngelNJP}. Hence it follows  that 
there  {\it must} exist  a degree of correlation  in the static energy variations in order to reproduce the experimental evidence  \cite{Engel2011b,EngelNJP}. Such correlation has been recently outlined as a process of paramount importance for understanding the signal arising from  macroscopic ensembles \cite{Engel2011d}.\\
Studies from first principles with molecular dynamics \cite{RengerPNAS}, suggest that the energy shift in individual BChls   in the FMO   are due to two contributions: 1) local, arising from different geometries of each BChl deformed within their fitting into the protein scaffold, 2) environmental, induced by the charge density of the protein due to its crystal structure. The latter process is generally accounted for by the charge density of $\alpha$-helices which extend across spatially extended areas involving in general, neighboring BChls. For example, BChla 3, 4 and 5, are situated close to the $\alpha$-helix 5 \cite{RengerPNAS}. Therefore,  the electric fields generated by the protein are created in spatially extended regions including several chromophores within a single monomer, and is expected that a set of neighboring chromophores will experience Stark shifts from conformational changes of a single $\alpha$ helix, i.e., a correlation on energy shifts.  If the shifts share  sign and approximate magnitude,  this situation will lead within the ensemble to small deviation from the average in the difference  $\tilde{\omega}_{\psi_2}-\tilde{\omega}_{\psi_1}$.

The fact that the main contributions in excitonic states $\ket{\psi_1}$ and $\ket{\psi_2}$ come from the neighboring sites 3, and 4 and 5 respectively, will imply in this scenario a degree of correlation in the energy of these excitonic states. 
Hence, FMO monomers with exciton energies  $\tilde{\omega}_{\psi_1}+\epsilon$ and $\tilde{\omega}_{\psi_2}+\epsilon^\prime$  fulfilling  $\langle{\epsilon^\prime}^2\rangle\simeq\langle\epsilon^2\rangle\approx \sigma_E^2$,  will result in $\epsilon\simeq \epsilon^{\prime}$ and $\langle(\epsilon-\epsilon^\prime)^2\rangle\ll\sigma_E$. With this assumption, is straightforward to average the signal according to the product of gaussian distributions $P(\epsilon)$, $P(\epsilon-\epsilon^\prime)$ which describe the correspondent variations of exciton 1 energy and of the relative difference among excitons 1 and 2 energies, having standard deviation $\sigma_\epsilon=\sigma_{E}$ and $\sigma_{\epsilon-\epsilon^\prime}\ll\sigma_{\epsilon}$. The result of the correlated averaging procedure is presented in  Fig.\ref{fig5}(a) (continuous line) which shows  an appreciable amplitude of oscillations in the signal at $t_2=2$ ps, in  good agreement with experimental evidence.  It is important to highlight that this model reproduces the inhomogeneous width recorded in absorption spectra or in the $\omega_1$ and $\omega_3$ dimension in 2D spectra, while concomitantly, it is able to reproduce  the enhanced lifetime of oscillation in the waiting time domain.\\
This last conclusion requires a more thorough analysis. Remember the Feynman diagrams Fig.\ref{fig4}(b), where the signal in the coherence time $t_1$ evolves at a frequency cycling among ground and excited electronic states, while that in the waiting time promotes oscillations among different excitonic ket-bra combinations. The local field will therefore lead in $t_1$ domain to a short dephasing decay time (as happens with the dashed line in Fig.\ref{fig5}(a), $\approx 100-200$ fs) proportional to the magnitude of the Stark shift due to the static fluctuations. The correlation  in the contributing chromophores of the excitons  allows  (as presented in Fig.\ref{fig5}(a)) the existence of one order of magnitude difference in the decay of the signal when compared to non-correlated inhomogeneity.   Hence, regardless of the magnitude of the Stark shift but sensitive to the gradient of the field producing it, the dephasing rate in the waiting time $t_2$ will decay slower than in $t_1$ when correlated fluctuations in the Stark shifts are present.  The extended local field induced Stark shift picture is further supported  by the recent experimental finding \cite{EngelNJP} that cross-peaks associated with excitonic states  whose main contributions involve neighboring chromophores have signals (in the waiting time domain) with smaller dephasing rates.  It cannot be explained based on heuristic arguments that  shifts should be correlated  by having equal signs at a given time. A more detailed  analysis based on the actual charge distribution over the $\alpha$ helices may address microscopically this possibility, but will not to be treated here.

The oscillation of the signal when correlated static inhomogeneities are present in  Fig.\ref{fig5}(a), resemble extremely well the behavior of the signal in the FMO 1-2 cross peak reported in \cite{Engel2011b,EngelNJP}. This has been achieved by doubling the Huang-Rhys factor of this representative vibration well within the bounds $0.12\le s<0.5$, $s=0.24$, and decreasing the homogeneous width of the mode $\gamma\rightarrow\gamma/8$ which is acceptable from the fact that fluorescence line narrowing or spectral hole burning already give an inhomogeneously broadened ensemble average of the vibrational manifolds. The lower bound $s=0.12$ has been used to infer the minimal effect of the mode of interest and keep consistent with experimental evidence.  Moreover, the discussion concerning the diagonal peaks enhancement due to inhomogeneity and phase shift remain valid for $s=0.24$. Quantitatively, the enhanced Huang-Rhys factor results in  an increased contrast of the vibronic bands compared to the ZPL. 
Experimentally, the modulation of electronic coherences has been  ascribed solely to overlap from peaks 1-2 and 1-3, which would produce a beating according to the recently estimated Hamiltonian \cite{Engel2011},  at a frequency of 38 cm$^{-1}$. However, this result highlights that such beating could arise solely from the discrete contribution of the most prominent vibronic mode. Hence, it must be underlined that many processes are observed at once in the 2D spectra and that beating of the amplitude in signals can arise from excitonic interference {\it and} electron-vibrational coupling.  Recently, resolution of the waiting time dynamics at different cross peaks has been achieved \cite{EngelNJP}, and therefore, the data is available to explore whether  an amplitude modulation having a frequency $\approx$ 36 cm$^{-1}$ is a general feature.
 
 In order to clarify  whether vibrational coherence is responsible for the long lasting oscillations present in the non-linear response of the FMO complex,   structural modifications were implemented \cite{Engel2011b}, whose effect was intended to change the vibrational manifold frequencies. Interestingly, none of the modifications   produced appreciable changes in the electronic coherence dephasing rate, nor  the oscillation frequency of the signal. Note however that our result implies a change in the period of the modulating envelope, and not that of the underlying  oscillations when structural modifications are generated. Moreover, the usual mechanism by which ensemble oscillations are damped from averaging dephased contributions, does not induce dephasing among signals regarding modes of different frequency. Fig.\ref{fig5}(b) shows that increasing the vibrational mode frequency inhomogeneities, will only decrease the beating depth without enhancing the dephasing rate of the signal. We model the signal  averaging according to a gaussian probability density function, with $\sigma_\omega$ of 10 cm$^{-1}$ in agreement with \cite{Small2000}.  This figure shows that using  a physically feasible Huang-Rhys factor is enough to  present  an appreciable modulation of the signal amplitude whose depth decreases as the vibrational mode inhomogeneities are larger.  Accordingly, static vibrational mode heterogeneities will lead to a decreased modulation of the signal amplitude but will not influence the overall decay of the signal.  In this way the robustness to structural modifications of the signal can be understood due to this novel {\it non-dephasing} effect over the ensemble inhomogenities.

\section{Conclusions and Perspectives} We proposed a theoretical model for the low energy band of the FMO complex comprising a vibronic discrete structure and study  the possibility to detect such vibronic dynamics with non-linear 2D optical spectroscopy. For experimentally determined parameters that characterize the electron-vibration coupling in FMO chromophores, the model is consistent with linear and non-linear optical spectra for the low energy band of the FMO complex. The main features of the vibrational side bands are variations of the signal phase  in the neighborhood of diagonal peaks and a modulation of the amplitude in the waiting time resolved signal at cross peaks, with a frequency corresponding to that of the energy of the vibrational discrete mode.  Complementarily, the robustness of excitonic coherence to dephasing can be further enhanced  by correlations in the static energy disorder within a single monomer, presumably  related with spatially extended electric field inhomogeneities that lead to correlated Stark shifts involving nearest neighbor chromophores.  

We show that correlated static disorder can lead to sample inhomogeneities that induce about an order of magnitude  enhancement of excitonic coherence lifetime as compared to the $t_1$ signal decay. When correlated static disorder is present, the discrete vibronic structure is not sensitive to ensemble inhomogeneities. Only under this conditions, the behavior of coherences at the single molecule level will  represent the ensemble averaged signal. Lastly, the resilience of the waiting time domain oscillations to structural modifications is understood in the 1-2 FMO cross peak as vibrations  generate a modulation in the signal amplitude, and not a variation in the signal intrinsic frequency. 
The signal presents a dichotomic sensitivity to excitonic and vibrational features, where the former are extremely sensitive to inhomogeneities while the latter are almost unaltered. Noticeable effects in the waiting time 2D signal are enhanced by both discreteness and coupling strength of the prominent vibrations.

The fact that the mode here studied might resonantly couple other electronic transitions in the FMO, and therefore, be useful to enhance delocalization among excitons, is a possibility that remains to be explored. Our study suggests that a detailed understanding of the vibrational environment, even at low frequencies, is crucial to achieve a full understanding of the electronic dynamics of pigment-protein complexes.

\section*{Acknowledgments}
 We acknowledge J. Caram, D. Hayes and G. S. Engel for providing access to experimental data and insightful discussions, and Financial support of the EU integrated project Q-ESSENCE, the EU STREP CORNER and the Alexander von Humboldt Foundation.
 

\end{document}